\documentclass[nofootinbib,twocolumn,prd]{revtex4}

\usepackage[dvips]{graphicx}
\usepackage{amsmath,amssymb,amsopn,bm,amsfonts}
\usepackage{color}
\usepackage{epstopdf}
\usepackage{subfigure}

\newcommand{\overbar}[1]{\mkern 1.5mu\overline{\mkern-1.5mu#1\mkern-1.5mu}\mkern 1.5mu}

\newcommand{\MeV}{\ensuremath{\mathrm{MeV}}}
\newcommand{\GeV}{\ensuremath{\mathrm{GeV}}}

\begin{document}
\title{Thermal photon emission from the $\pi \rho \omega$ system}

\author{Nathan P. M. Holt}
\email{nathan.holt@physics.tamu.edu}
\affiliation{Cyclotron Institute and Department of Physics\,\&\,Astronomy,
Texas A{\&}M University, College Station, Texas 77843-3366, USA }

\author{Paul M. Hohler}
\email{pmhohler@comp.tamu.edu}
\affiliation{Cyclotron Institute and Department of Physics\,\&\,Astronomy,
Texas A{\&}M University, College Station, Texas 77843-3366, USA }

\author{Ralf Rapp}
\email{rapp@comp.tamu.edu}
\affiliation{Cyclotron Institute and Department of Physics\,\&\,Astronomy,
Texas A{\&}M University, College Station, Texas 77843-3366, USA }

\begin{abstract}
We investigate thermal photon emission rates in hot hadronic matter
from a system consisting of $\pi$, $\rho$, and $\omega$ mesons. The rates are
calculated using both relativistic kinetic theory with Born diagrams as well as
thermal field theory at the two-loop level. This enables us to cross-check our
calculations and to manage a pole contribution that arises in the Born approximation
corresponding to the $\omega \to \pi^0 \gamma$
radiative decay. After implementing hadronic
form factors to account for finite-size corrections, we find that the resulting
photo-emission rates are comparable to existing results from $\pi\rho\to\pi\gamma$
processes in the energy regime of 1--3 GeV. We expect that our new sources will
provide a non-negligible contribution to the total hadronic rates, thereby enhancing
calculated thermal photon spectra from heavy-ion collisions, which could improve
the description of current direct-photon data from experiment.
\end{abstract}

\maketitle

\section{Introduction}
\label{sec:intro}

One of the goals at the forefront of nuclear physics is to understand the phase
diagram of strongly interacting matter~\cite{Akiba:2015jwa,Heinz:2015tua}.  At
sufficiently high temperatures, hadronic matter undergoes
a transition from color-neutral particles to a quark-gluon plasma (QGP) with
deconfined color charge. As this transition is believed to be a crossover, the
properties of hadronic matter are expected to change continuously as
temperatures approach the pseudo-critical one,
$T_{\rm pc} \approx 160$\,MeV~\cite{Borsanyi:2010bp,Bazavov:2011nk}.
The phases of QCD matter can be explored in ultra-relativistic heavy-ion
collisions (URHICs) where a fireball of high-temperature strongly-interacting
matter is created. This matter rapidly cools and ultimately decouples into
confined particles
which are then detected. Hadronic probes of the medium undergo strong
rescattering and alterations while traversing the fireball thus losing
direct information about the properties of the hot and dense phases of
the evolution. Electromagnetic (EM) probes (photons and dileptons), on the
other hand, are emitted throughout the evolution of the fireball, escaping
relatively unaltered since their mean free path is much longer than the size
of the fireball. They thus encode information on properties of the fireball
which are not easily accessible otherwise, e.g., on its interior temperatures,
the total space-time volume, and evolution history of its collective properties
(see, e.g., the reviews in
Refs.~\cite{Rapp:2009yu,Gale:2009gc,Alam:2015cja} for further information
on the role of EM probes in URHICs).

Theoretical models of EM probes in URHICs relate dilepton and photon spectra
to the properties of the emitting medium. The calculated emission spectra
of dileptons~\cite{Rapp:2009yu,Rapp:2013nxa} agree well with experimental
measurements thus far~\cite{Arnaldi:2008fw,Adamova:2006nu,Geurts:2012rv},
while similar calculations of the photon spectra~\cite{vanHees:2011vb} indicate
potential discrepancies with recent measurements of direct
photons~\cite{Adare:2008fq,Adare:2011zr,Wilde:2012wc,Lohner:2012ct,Yang:2014mla}.
This discrepancy concerns both the spectral yields and the photon elliptic flow
($v_2$).

Much theoretical effort has gone into addressing this putative
``puzzle"~\cite{Liu:2009kta,vanHees:2011vb,Holopainen:2011pd,Dion:2011pp,
Mohanty:2011fp,Shen:2013vja,Linnyk:2013wma,Bzdak:2012fr,vanHees:2014ida,
Monnai:2014kqa,Gale:2014dfa,McLerran:2015mda}.
The tentative conclusion of most of these approaches is that there are
currently unaccounted-for thermal photon sources from the strongly-interacting
medium.  The difficulty in the problem and the diversity in possible solutions
comes from identifying these new sources. For example, additional sources
from a hot QGP, which contribute early in the evolution, are disfavored as
they carry a small $v_2$ (the fireball needs a time of the order of the
nuclear size, $\tau_{\rm FB}\sim R_A$, to build up the momentum anisotropy)
and too hard of a spectral slope. This leads one to consider sources which
contribute later in the fireball evolution as prime candidates for generating
the necessary $v_2$~\cite{vanHees:2011vb,Shen:2013vja,Linnyk:2013wma,vanHees:2014ida}.

The present paper aims at identifying such sources by investigating
novel sources of photon production from hot hadronic matter, and in this
way contribute to a more complete characterization of the electromagnetic
emissivity of QCD matter in a regime of moderate temperatures.
Specifically, we will explore photon emission rates from a thermally
equilibrated system composed of $\pi$, $\rho$, and $\omega$ mesons, and put
their relevance into context with existing rate calculations (alternatively
our calculated scattering matrix elements can be used in transport
simulations where the momentum distributions are not necessarily thermal).
The coupling strength of the $\pi\rho\omega$ vertex, which is pivotal to our
analysis, is known to be large~\cite{Wess:1971yu,Witten:1983tx}.
It was instrumental in identifying the $\omega$ $t$-channel exchange
in the $\pi \rho \to \pi \gamma$ process as an important contributor to
the photon emissivity of a hot meson gas~\cite{Turbide:2003si} (and, in
fact, to the $\omega \to \pi^0 \gamma$ decay~\cite{Rapp:1999qu}).
Thus, a more systematic analysis of pertinent processes, including the
$\omega$ meson as an external particle (incoming or outgoing), is warranted.
Since the $\omega$ is not a stable particle under strong interactions
(it can decay into $\pi\rho$ or $3\pi$), some care has to be taken when
evaluating pertinent scattering diagrams. To ensure that this is done
correctly, we calculate the emission rates using both kinetic theory and
thermal field theory techniques to identify potential issues and cross-check
the results.

The remainder of this paper is organized as follows.
In Sec.~\ref{sec:photons}, we briefly recapitulate two methods for
calculating thermal photon production in the context of the $\pi \rho \omega$
system, including the specification of the interaction Lagrangians.
In Sec.~\ref{sec:kt}, we calculate photo-emission rates using
kinetic theory (KT) with Born scattering amplitudes.
In Sec.~\ref{sec:tft}, we use thermal field theory (TFT) to compute
two-loop diagrams of the photon self-energy to both check the KT results
and address some intricacies in the kinetic approach.
In Sec.~\ref{sec:results}, we present our results and compare them
to existing hadronic photo-emission rates.
We conclude in Sec.~\ref{sec:conclusion}.

\section{Thermal Photon Emission Rates}
\label{sec:photons}

The thermal photon production rate can be written as~\cite{McLerran:1984ay}
\begin{equation}
q_0 \frac{dR_{\gamma}}{d^3 q} = - \frac{\alpha_{\mathrm{em}}}{\pi^2}
f(q_0,T) \, \mathrm{Im} \, \Pi^T_{\mathrm{em}}(q_0= |\vec{q}\,|,T) \ ,
\label{eq:rate}
\end{equation}
where $q_0$ and $|\vec{q}\,|$ are the photon's energy and
three-momentum magnitude, respectively, and $f$ is the
thermal Bose-Einstein distribution function.
Employing the vector meson dominance (VMD) model~\cite{Sakurai}, the
3D-transverse part of the EM current-current correlator,
$\Pi^T_{\mathrm{em}}$, is proportional to the in-medium vector
meson spectral function (dominated by the $\rho$ meson) via
\begin{equation}
\Pi^T_{\mathrm{em}}(q_0, \vec{q}\,) = \frac{ ( m_{\rho}^{(0)} )^4 }{g_{\rho}^2}
 \, D^T_{\rho}(q_0,\vec{q}\,) \ ,
\label{eq:VMD}
\end{equation}
with a dressed $\rho$ propagator
\begin{equation}
D^T_\rho(q_0,\vec{q}\,) = \left[q_0^2 - |\vec{q}\,|^2 - ( m_\rho^{(0)} )^2
-\Sigma^T_\rho(q_0,\vec{q}\,)\right]^{-1} \ ,
\end{equation}
where $m_\rho^{(0)}$ is the bare $\rho$ mass and $\Sigma_{\rho}^T$ is the
in-medium $\rho$ self-energy.
In the VMD model, the thermal photon emission calculation thus amounts
to calculating the $\rho$ meson self-energy, usually done using TFT.

Alternatively, calculations of thermal photon emission can be done using KT.
In Ref.~\cite{Weldon:1983jn}, it was shown that the imaginary
parts of self-energies can be expressed as vacuum scattering amplitudes
(i.e. using zero-width external particles) folded with appropriate thermal
statistical weightings and integrated over the pertinent phase space.
In this framework, the thermal photon emission rate is expressed in
terms of a $1+2 \to 3+\gamma$ scattering process given by
\begin{equation}
\begin{split}
\label{eq:KTrate}
q_0 \frac{dR_{\gamma}}{d^3 q} & = \mathcal{N} \int
\frac{d^3 p_1}{(2 \pi)^3 2E_1} \frac{d^3 p_2}{(2 \pi)^3 2E_2}
\frac{d^3 p_3}{(2 \pi)^3 2E_3} \\ & \times (2 \pi)^4 \delta^4
(p_1+p_2-p_3-q) \, \overbar{|\mathcal{M}|^2} \\ & \times f(E_1,T)
f(E_2,T) \frac{[1 + f(E_3,T)]}{2(2 \pi)^3} \ ,
\end{split}
\end{equation}
where $\mathcal{N}$ is the combined spin and isospin
degeneracy of the incoming particles and $\overbar{|\mathcal{M}|^2}$
is the initial-state spin- and isospin-averaged scattering amplitude of the
process under consideration.

The microscopic ingredients to the photo-emission rates using the above
frameworks are the $\rho$ self-energy (for TFT) and photon-producing
scattering amplitudes (for KT).
Both quantities can be derived on the same footing from an underlying
interaction Lagrangian. We start with a Lagrangian for free $\pi$ and $\rho$
fields,
\begin{equation}
\label{eq:fields}
\mathcal{L}_{\pi} + \mathcal{L}_{\rho} =
\frac{1}{2}\partial_{\mu} \vec{\pi} \cdot \partial^{\mu}
\vec{\pi} - \frac{1}{2} m_{\pi}^{2} \vec{\pi} \cdot \vec{\pi}
- \frac{1}{4} \vec{\rho}_{\mu \nu} \cdot \vec{\rho}^{\,\mu \nu}
+ \frac{1}{2} m_{\rho}^{2} \vec{\rho}_{\mu} \cdot \vec{\rho}^{\, \mu} \ ,
\end{equation}
with the usual definition of the $\rho$ field strength tensor as
\begin{equation}
\vec{\rho}_{\mu \nu} = \partial_{\mu} \vec{\rho}_{\nu} -
\partial_{\nu} \vec{\rho}_{\mu} \ .
\end{equation}
Interactions between the $\pi$ and $\rho$ mesons are generated
by minimally coupling derivative terms of the $\pi$ and $\rho$
fields to a $\rho$ gauge field~\cite{Kroll:1967it},
\begin{equation}
\label{eq:gauge}
\begin{split}
\partial_{\mu} &
\to \partial_{\mu} + i g_{\rho} \vec{\rho}_{\mu} \cdot \vec{T}
\\
\partial_{\mu} \pi^a &
\to \partial_{\mu} \pi^a + g_{\rho} \epsilon_{abc} \rho_{\mu}^b  \pi^c
\\
\partial_{\mu} \rho_{\nu}^a &
\to \partial_{\mu} \rho_{\nu}^a + g_{\rho}\epsilon_{abc}\rho_{\mu}^b\rho_{\nu}^c \ ,
\end{split}
\end{equation}
where $\vec{T}$ is the isospin-1 operator
with components of $(T^a)_{bc} = -i \epsilon_{abc}$~\cite{Ericson:1988gk}
and $g_{\rho}$ is the $\rho\pi\pi$ coupling constant.  Introducing the
above covariant derivatives into the free-field Lagrangian results
in $\rho \pi \pi$ and $\rho \rho \rho$ interactions of
\begin{eqnarray}
\label{eq:interaction1}
\mathcal{L}_{\rho \pi \pi} = -g_{\rho} \vec{\rho}^{\,\mu} \cdot
(\partial_{\mu} \vec{\pi} \times \vec{\pi}) \ , \\
\nonumber\\
\mathcal{L}_{\rho \rho \rho} = -\frac{1}{2} g_{\rho}
\vec{\rho}^{\,\mu \nu} \cdot (\vec{\rho}_{\mu} \times \vec{\rho}_{\nu}) \ .
\end{eqnarray}
Interactions with the isosinglet vector $\omega$ field are included via
a Wess-Zumino term~\cite{Wess:1971yu,Witten:1983tx},
\begin{equation}
\label{eq:interaction2}
\mathcal{L}_{WZ} = g_{\pi\rho\omega} \epsilon^{\mu \nu \alpha \beta}
\partial_{\alpha} \omega_{\beta} \partial_{\mu} \vec{\rho}_{\nu}
\cdot \vec{\pi} \ .
\end{equation}
Applying the gauging procedure of Eq.~(\ref{eq:gauge}) to the
Wess-Zumino term generates a contact term
\begin{equation}
\label{eq:interaction3}
\mathcal{L}_{\rho \rho \pi \omega} = g_{\pi\rho\omega} g_{\rho}
\epsilon^{\mu \nu \alpha \beta} \partial_{\alpha} \omega_{\beta}
(\vec{\rho}_{\mu} \times \vec{\rho}_{\nu}) \cdot \vec{\pi} \ .
\end{equation}

Interactions with photons are introduced through VMD.
We neglect the coupling of the $\omega$ field to the electromagnetic
current, as the $\omega \gamma$ coupling is suppressed by a factor of
$\approx$3--4 relative to the $\rho \gamma$ coupling~\cite{Sakurai}.
Following Ref.~\cite{Turbide:2003si}, we parametrize the $\rho \gamma$
coupling with $C_{\rho}$,
\begin{equation}
\mathcal{L}_{EM} = -A^{\mu} C_{\rho} m_{\rho}^2 \rho^{0}_{\mu} \ ,
\label{eq:EM-coupling}
\end{equation}
where $A_{\mu}$ is the photon field and $\rho^{0}_{\mu}$ is the
neutral $\rho$ field.

For a realistic description of interaction processes, we also need to account
for the finite size of the mesons, which we do in the standard way by
introducing previously determined dipole hadronic form factors at each
vertex~\cite{Rapp:1999qu,Turbide:2003si}.
For $s$-channel decay processes, we employ at each vertex the form
\begin{equation}
F(s) = \left(\frac{2 \Lambda^2 + m_R^2}{2 \Lambda^2 + \left[E_2(p_{CM})
+ E_3(p_{CM})\right]^2} \right)
\label{eq:FFs}
\end{equation}
where $E_i(p_{CM})=\sqrt{m_i^2 + p_{CM}^2}$ and $p_{CM}(s)$ is the
center-of-mass momentum of each hadronic decay particle, $i$=2,3,
$\Lambda$ is an hadronic cut-off parameter (taken to be 1 GeV in the
present work), and $m_R$ is the mass of the resonant (or decay) particle.
For $t$-channel scattering processes, we employ at each vertex the form
\begin{equation}
F(t) = \left( \frac{2\Lambda^2}{2\Lambda^2 + |t|} \right)^2 \ ,
\end{equation}
with $t=(p_1-p_3)^2$ for incoming ($p_1$) and outgoing ($p_3$) 4-momenta
(and likewise for $u$-channel processes).

Save for $s$-channel decay processes or those processes with
individually gauge-invariant vertices,
the above implementation of form factors renders it a rather involved task
to maintain gauge invariance, especially for scattering processes where
multiple diagrams contribute~\cite{Kapusta:1991qp}. Therefore, we follow
the simplified prescription of Ref.~\cite{Turbide:2003si} by introducing an
overall factorized form factor squared for each scattering matrix element
with an appropriately defined average momentum transfer.
This is done by identifying the dominant diagram in the
interaction process, i.e., the diagram with the largest contribution to the
high-energy photo-emission rate of a given process (at low energies the form factor
effects are small). Typically, this will be a $t$-channel exchange diagram with
the lightest exchange particle (e.g., pion exchange in $\pi\rho\to\gamma\pi$),
since $s$-channel processes are suppressed by propagators of
the form $(s-m_R^2)^{-1}$. The average momentum transfer, $\overbar{t}$, is evaluated from
 \begin{multline}
\label{eq:tbar}
\left( \frac{1}{m_X^2 + |\overbar{t}|} \right)^2 \left(
\frac{2\Lambda^2}{2\Lambda^2 + |\overbar{t}|} \right)^8 \\ =
- \frac{1}{4 q_0^2} \int_{0}^{-4 q_0^2} dt \left( \frac{1}{m_X^2 +
|t|} \right)^2 \left( \frac{2\Lambda^2}{2\Lambda^2 + |t|} \right)^8 \ ,
\end{multline}
(and likewise if a $u$-channel process dominates), and is a
function of only the photon energy and exchange-particle mass
in the dominant diagram.
This enables us to factorize the form factor from the total amplitude,
\begin{equation}
\overbar{|\mathcal{M}|^2} = \overbar{|\mathcal{M_{\rm point}}|^2}
F(\overbar{t})^4 \ ,
\end{equation}
and thus retain the gauge invariance in the amplitude, $\mathcal{M}_{\rm point}$,
which is evaluated for point-like vertices.

The coupling constants $g_{\rho}$ and $g_{\pi\rho\omega}$ are evaluated using
data from $\rho \to \pi \pi$ and
$\omega \to \pi^0 \gamma$ decays. In principle, the couplings $C_{\rho}$
and $g_{\rho}$ are related via VMD, such that $C_{\rho} = e/g_{\rho}$.
However, for our Born diagrams, where we are using zero-width $\rho$ mesons,
we will instead use experimental data from the $\rho \to e^{+} e^{-}$ decay
to evaluate $C_{\rho}$. Using the $s$-channel form factor of Eq.~(\ref{eq:FFs}),
the decay rate becomes
\begin{equation}
\Gamma_{1 \to 2+3} = \frac{p_{CM} \overbar{|\mathcal{M}|^2}
F^2(p_{CM})}{8 \pi m_1^2} \ ,
\end{equation}
where  $m_1$ is mass of the decaying particle and
$\overbar{|\mathcal{M}|^2}$ is the initial-state averaged and
final-state summed matrix element of the associated decay process.
Using Particle Data Group~\cite{Agashe:2014kda} values for
$\Gamma_{\rho \to \pi \pi} = 149.1$ $\MeV$,
$\Gamma_{\omega \to \pi^0 \gamma} = 0.703$ $\MeV$, and
$\Gamma_{\rho \to e^{+} e^{-}} = 7.04$ $\mathrm{keV}$,
and a form factor cutoff of $\Lambda = 1$ $\GeV$, we find coupling constants
of $g_{\rho} = 5.98$, $g_{\pi\rho\omega} = 21.6\,\GeV^{-1}$\footnote{This
slightly differs from the value of 25.8\,$\GeV^{-1}$ in Ref.~\cite{Rapp:1999qu}
since there the VMD value of $C_\rho=e/g=0.052$ is used.},
and $C_{\rho} = 0.0611$.

Having established all necessary interaction vertices, including form factor
effects and the quantitative evaluation of all coupling constants, we now
proceed to the photo-emission rate calculations.

\section{Kinetic Theory}
\label{sec:kt}

\begin{figure*}[t!]
\centering
\subfigure[\, $s$-channel]{
\includegraphics[scale=0.4]{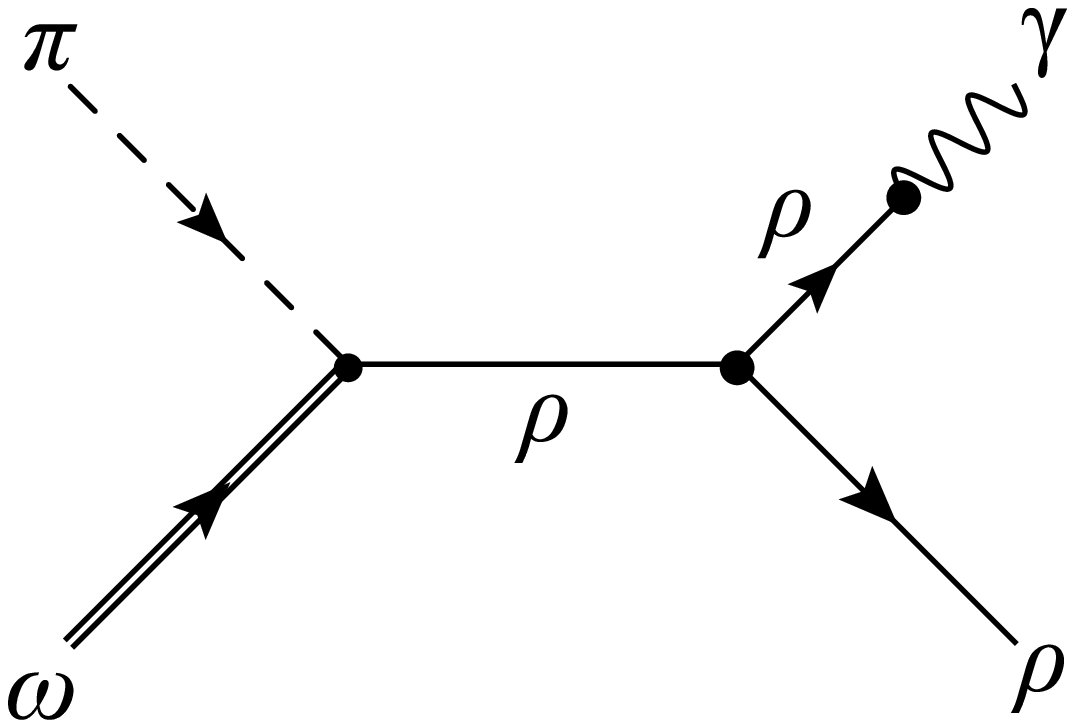}}
\subfigure[\, contact term]{
\includegraphics[scale=0.4]{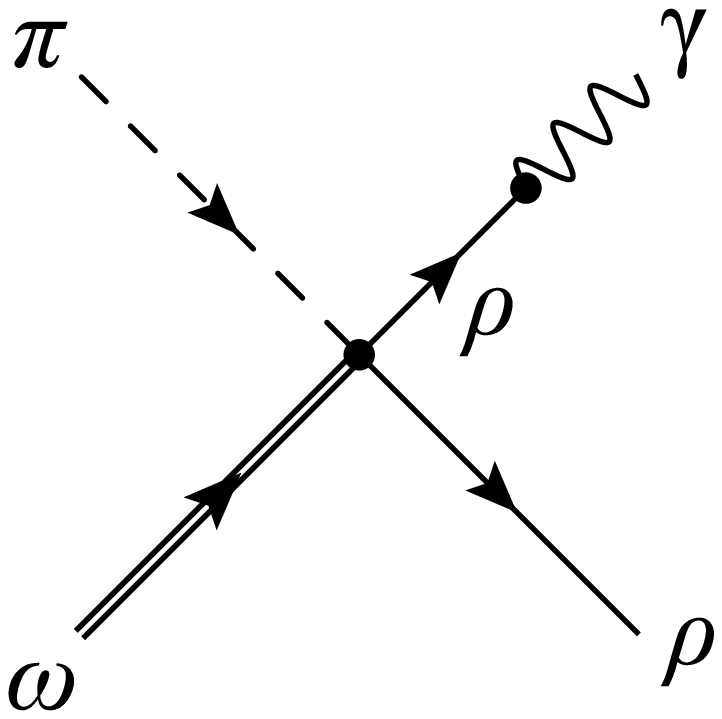}}
\subfigure[\, $t$-channel]{
\includegraphics[scale=0.4]{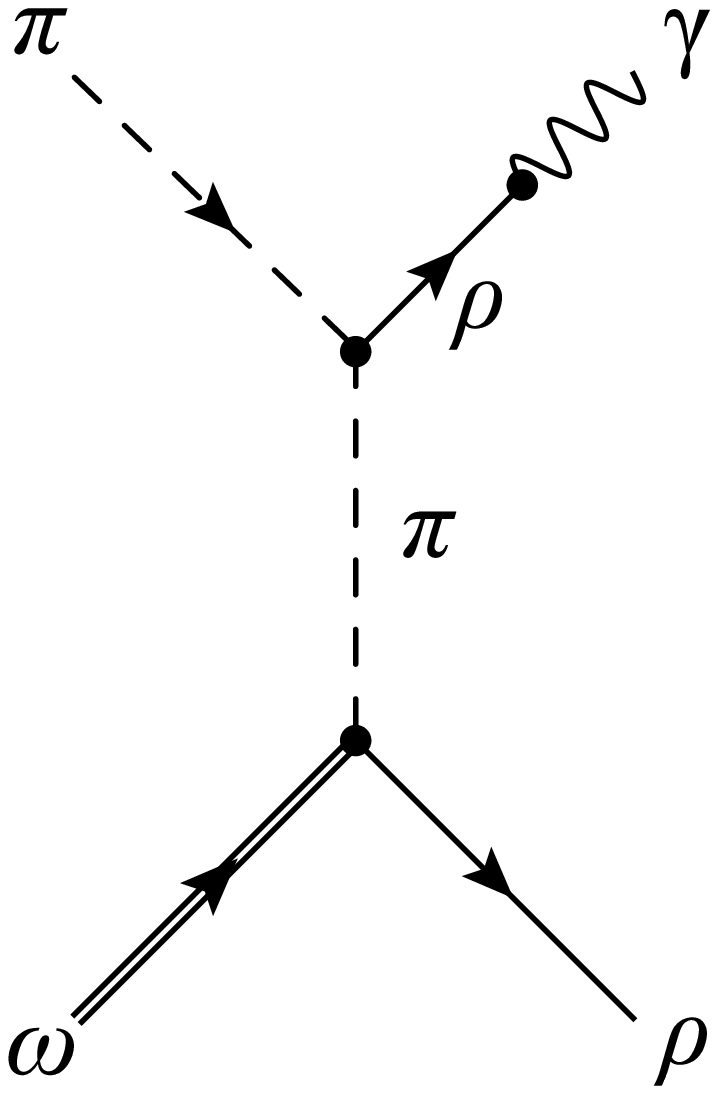}}
\subfigure[\, $u$-channel]{
\includegraphics[scale=0.4]{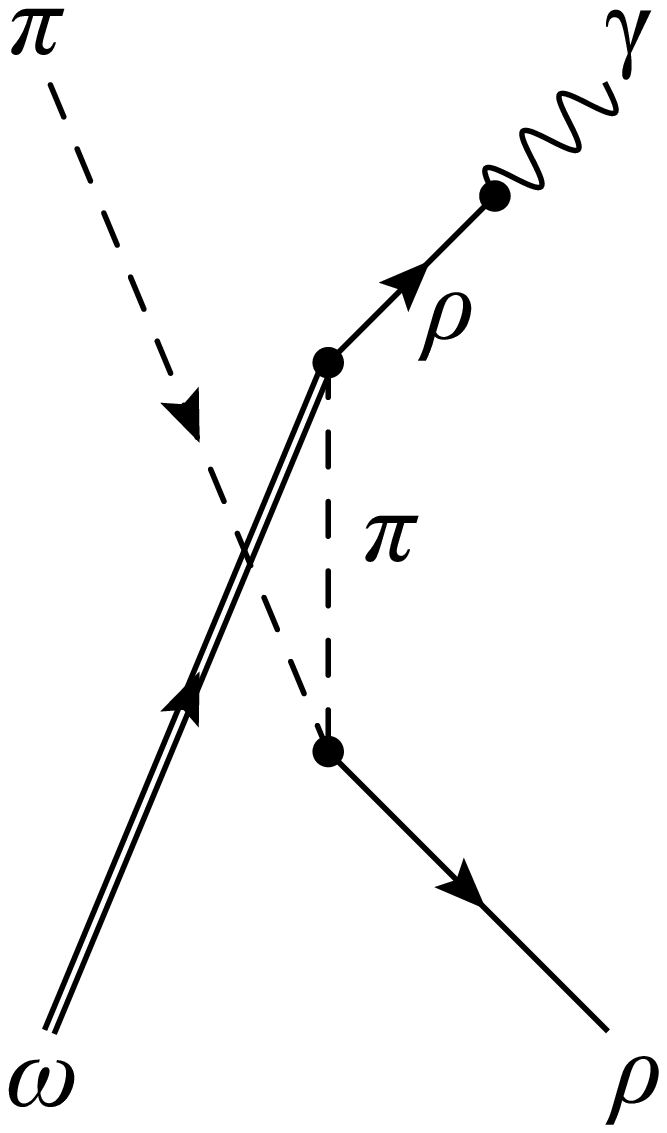}}
\subfigure[\, $s$-channel]{
\includegraphics[scale=0.4]{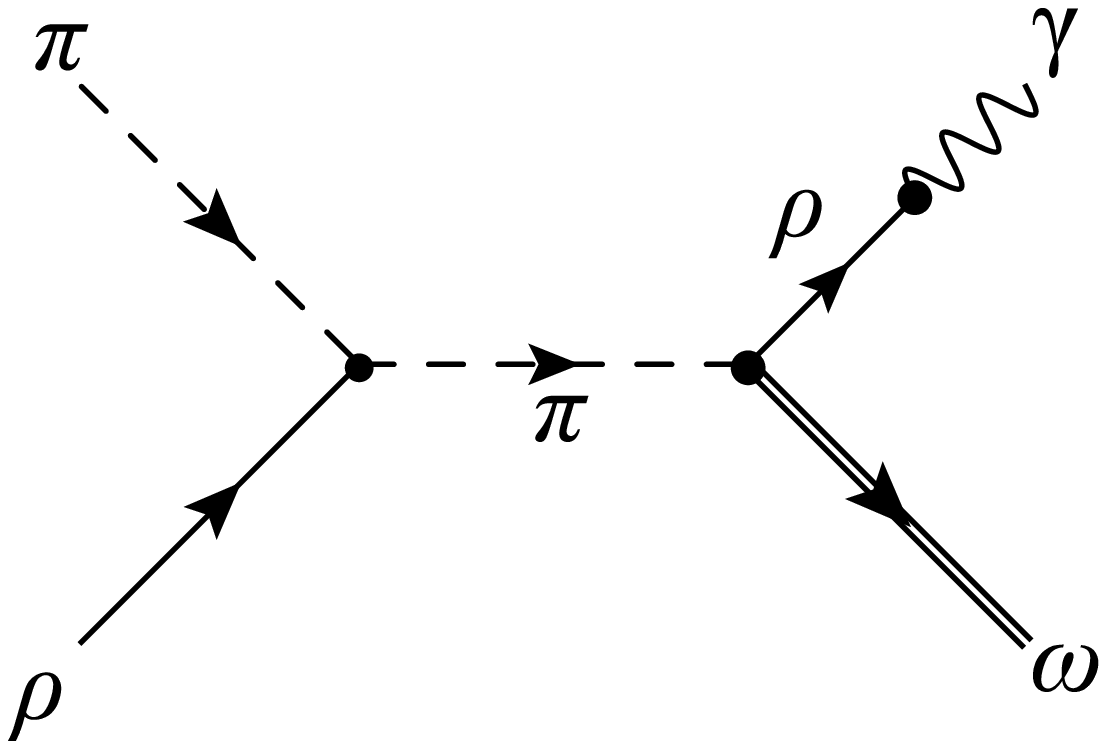}}
\subfigure[\, contact term]{
\includegraphics[scale=0.4]{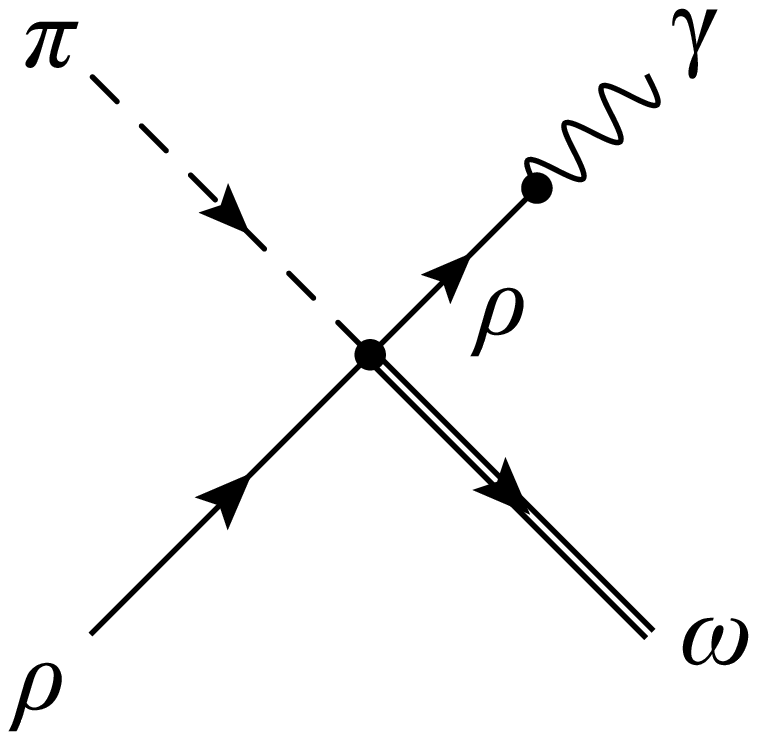}}
\subfigure[\, $t$-channel]{
\includegraphics[scale=0.4]{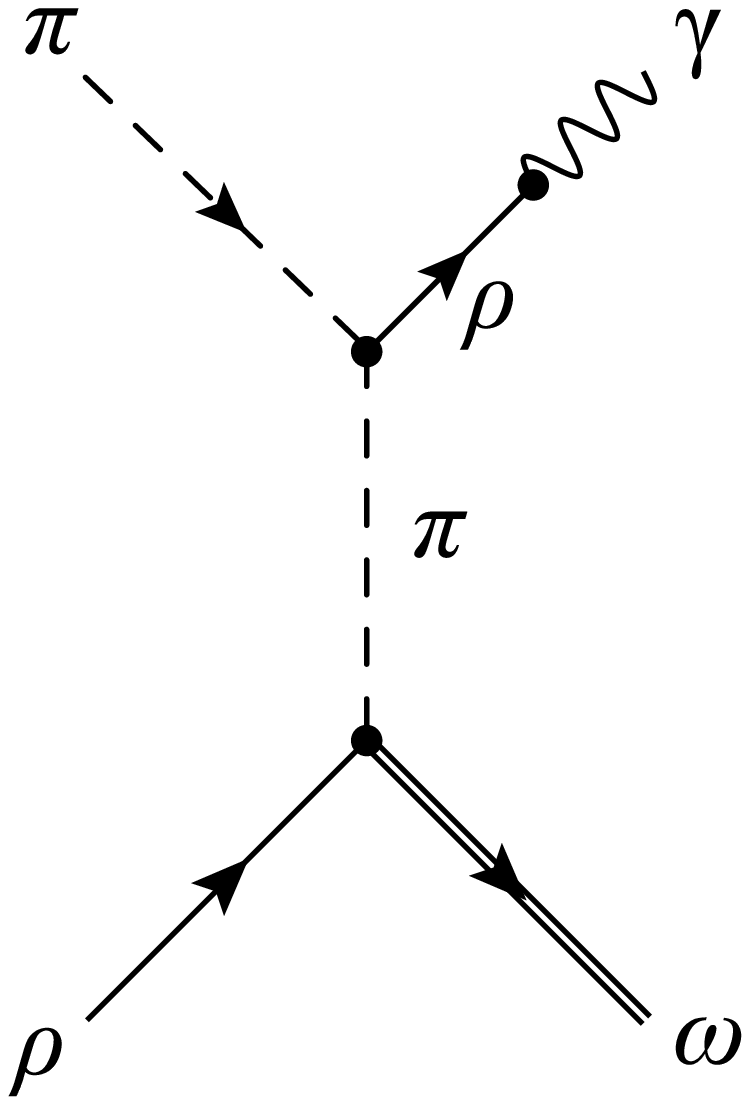}}
\subfigure[\, $u$-channel]{
\includegraphics[scale=0.4]{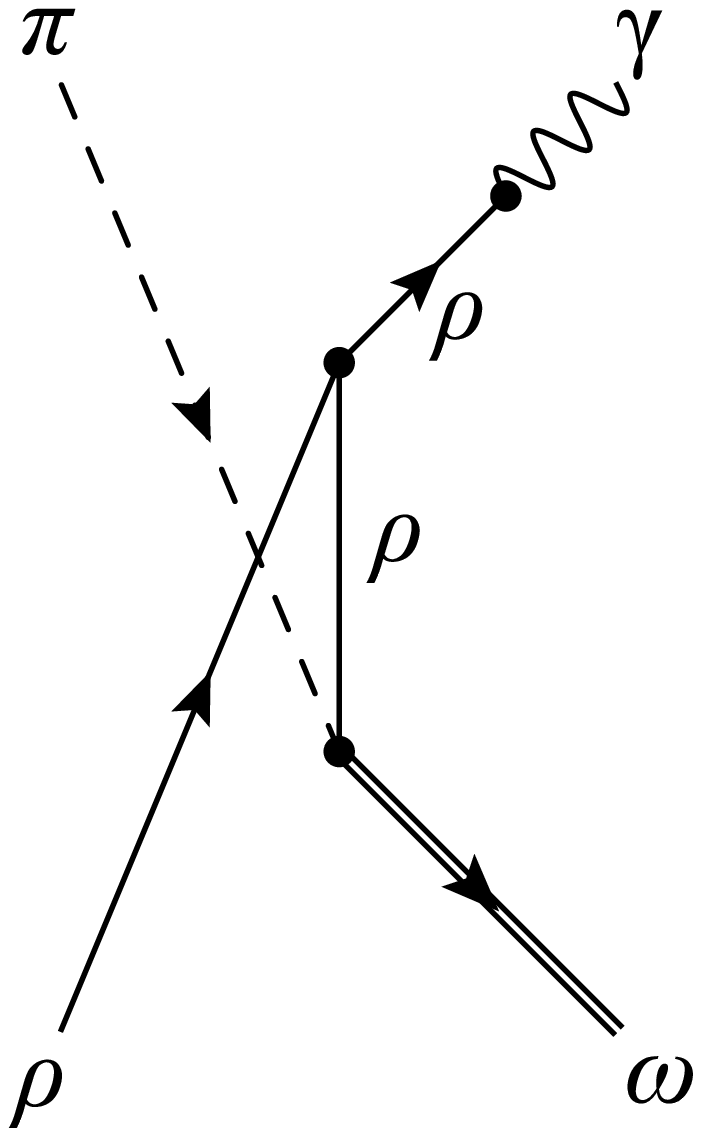}}
\subfigure[\, $s$-channel]{
\includegraphics[scale=0.4]{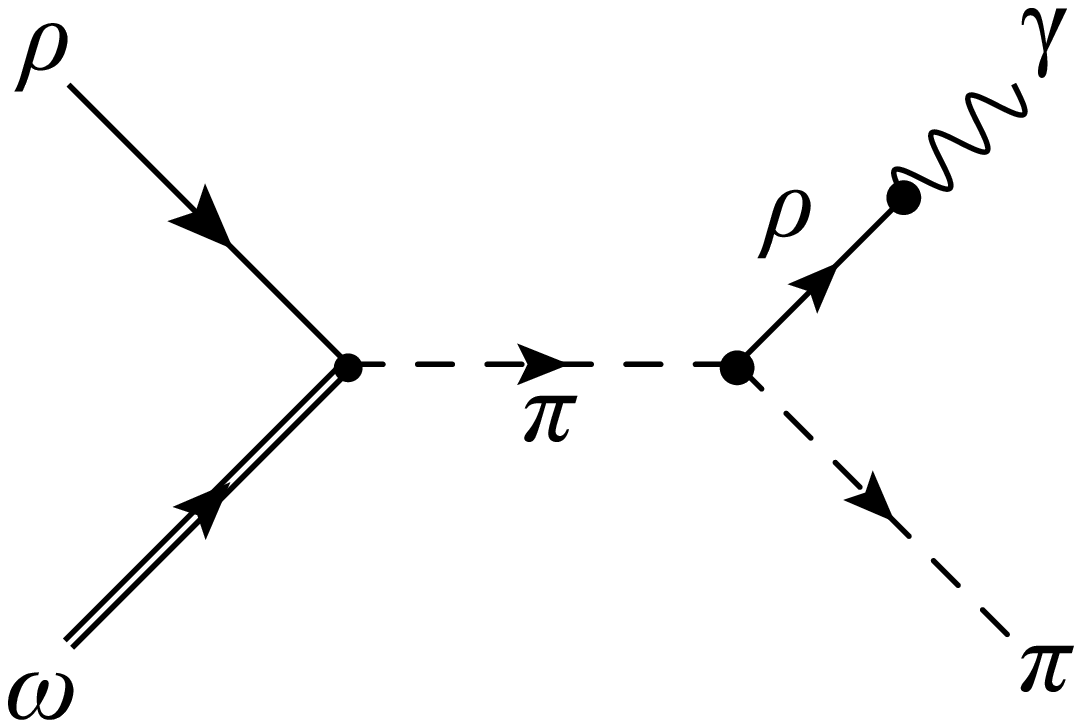}}
\subfigure[\, contact term]{
\includegraphics[scale=0.4]{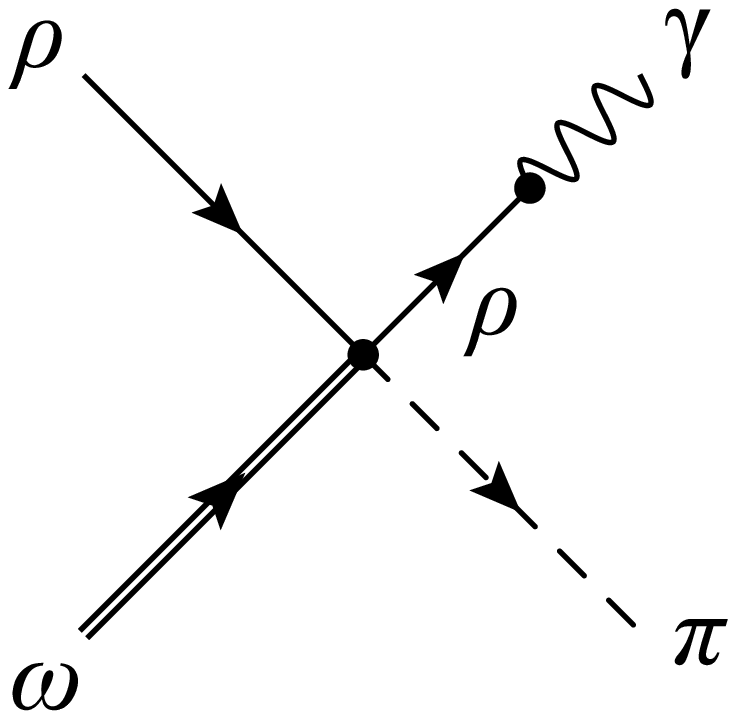}}
\subfigure[\, $t$-channel]{
\includegraphics[scale=0.4]{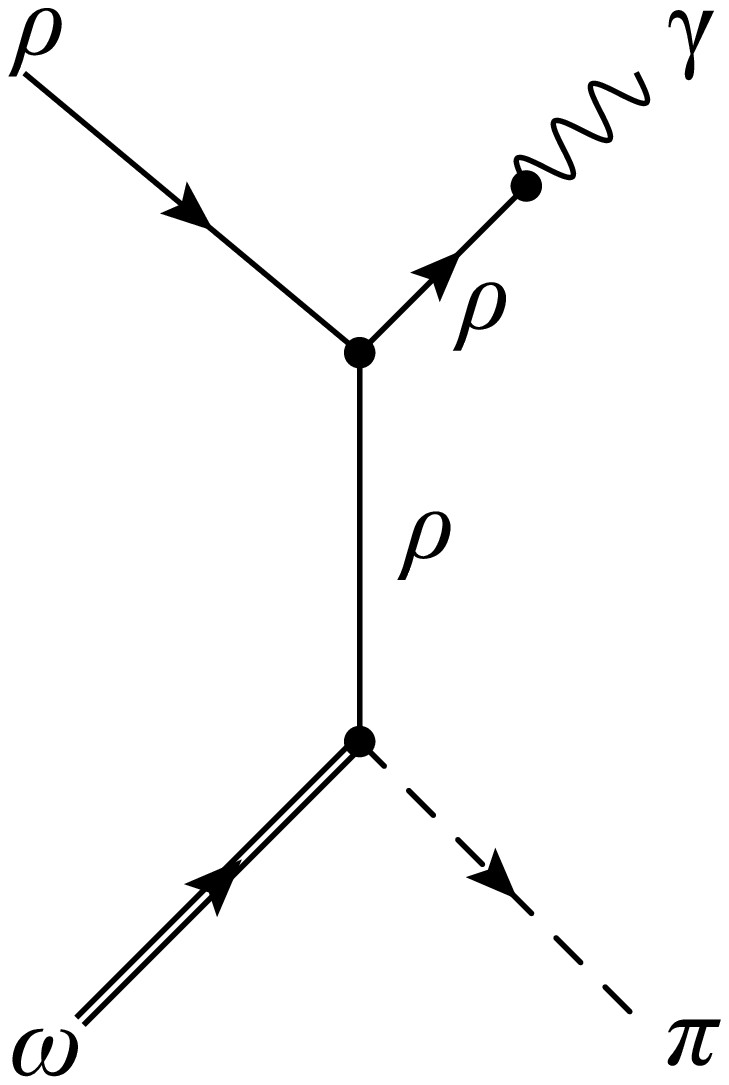}}
\subfigure[\, $u$-channel]{
\includegraphics[scale=0.4]{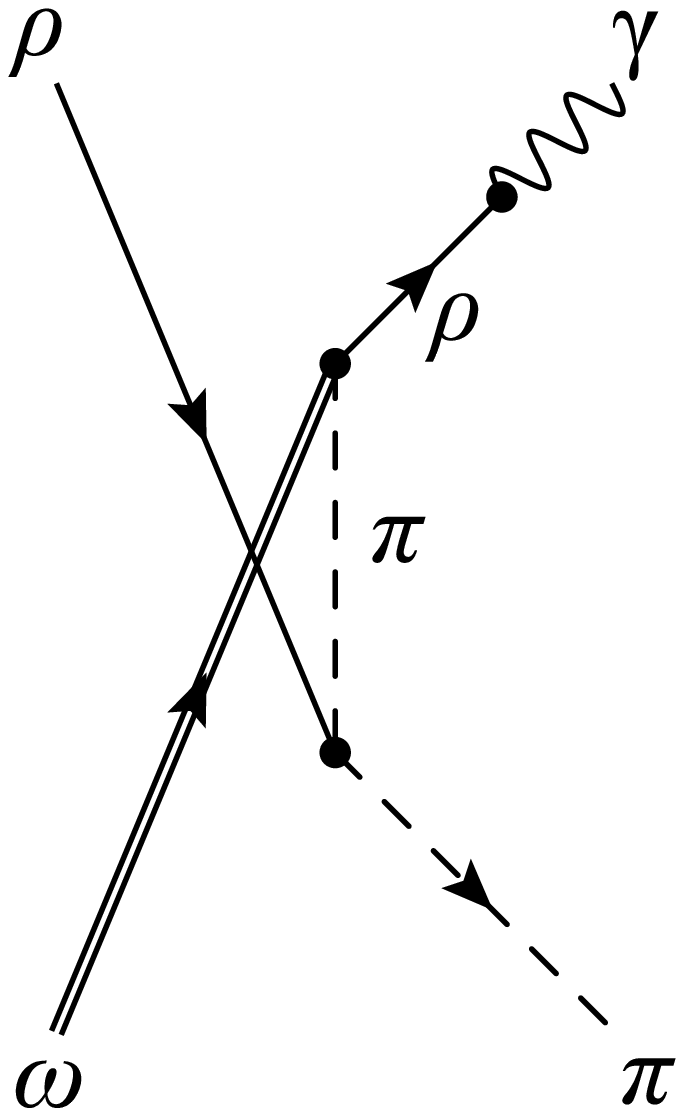}}
\caption{Feynman Born diagrams for photon emission from the
$\pi \rho \omega$ system, i.e., Figs. (a)-(d) for
$\pi \omega \to \gamma \rho$, Figs. (e)-(h) for
$\pi \rho \to \gamma \omega$, and Figs. (i)-(l) for
$\rho \omega \to \gamma \pi$. }
\label{fig:diagrams}
\end{figure*}

Within the framework of relativistic KT, thermal photon emission from the
$\pi \rho \omega$ system with external $\omega$ particles arises
from three 2$\to$2 scattering processes: $\pi \rho \to \gamma \omega$,
$\pi \omega \to \gamma \rho$, and $\rho \omega \to \gamma \pi$.  Each
of the three processes is composed of $s$-, $t$-, $u$-channel diagrams
and a contact ($c$) term to ensure gauge invariance.  The diagrams
comprising each process in the $\pi \rho \omega$ system are shown in
Fig.~\ref{fig:diagrams}.

The evaluation of photo-emission rates using Eq.~(\ref{eq:KTrate})
requires the coherent sum of the four amplitudes of each diagram;
\begin{equation}
|\mathcal{M}|^2 = |\mathcal{M}_s
+\mathcal{M}_t+\mathcal{M}_u+\mathcal{M}_c|^2 \ .
\end{equation}
We calculate the amplitudes by applying Feynman rules to the diagrams
shown in Fig.~\ref{fig:diagrams} and using the Lagrangian interactions
and form factor procedure detailed in Sec.~\ref{sec:photons}. The explicit
expressions for the matrix elements are given in
Appendix~\ref{app:amplitudes}.

Straightforward calculations of rates for the $\pi \rho \omega$
system using KT can be done for
the $\pi \rho \to \gamma \omega$ and $\rho \omega \to \gamma \pi$
processes, while the $\pi \omega \to \gamma \rho$ process
reveals a subtlety. In the $u$-channel diagram, Fig.~\ref{fig:diagrams}(d),
the exchanged pion can go on-shell, such that
$u = (p_{\omega}-p_{\gamma})^2 = m_{\pi}^2$.  This induces a non-integrable
singularity in the phase space of the rate calculation using
Eq.~(\ref{eq:KTrate}). This pion pole configuration corresponds to the
$\omega \to \pi^0 \gamma$ radiative decay, which, in fact, has already
been included in previous photo-emission
calculations~\cite{Rapp:1999qu,Turbide:2003si}.
Therefore we must eliminate this contribution from our results to avoid
double-counting of the radiative $\omega$ decay.
This is facilitated by the structure of the Wess-Zumino term,
Eq.~(\ref{eq:interaction2}), which renders a diagram individually
gauge-invariant when an outgoing $\rho$ meson is converted to a photon.
This feature allows us to separate the $u$-channel from
the other three diagrams in the $\pi \omega \to \gamma \rho$ process while
maintaining gauge invariance. Na\"{i}vely, we can avoid the $\omega$ decay
by excluding timelike pion configurations with $u >0$
from the integration range in Eq.~(\ref{eq:KTrate}), but \textit{a priori}
this is not a rigorous justification for this choice.
To scrutinize this prescription to avoid double-counting of
the $\omega$ decay, we turn to TFT where such an ambiguity does
not occur.

\section{Thermal Field Theory}
\label{sec:tft}

Thermal field theory provides a rigorous framework for the
calculation of photo-emission rates. As discussed above, within the
VMD model the relevant quantity is the $\rho$ meson self-energy; recall
Eqs.~(\ref{eq:rate}) and (\ref{eq:VMD}).
Each process and channel considered in the KT calculation has a
corresponding $\rho$ self-energy diagram associated with it.

\begin{figure}[!h]
\centering
\subfigure[\, $\rho \omega \to \gamma \pi$ process]{
\includegraphics[scale=0.5]{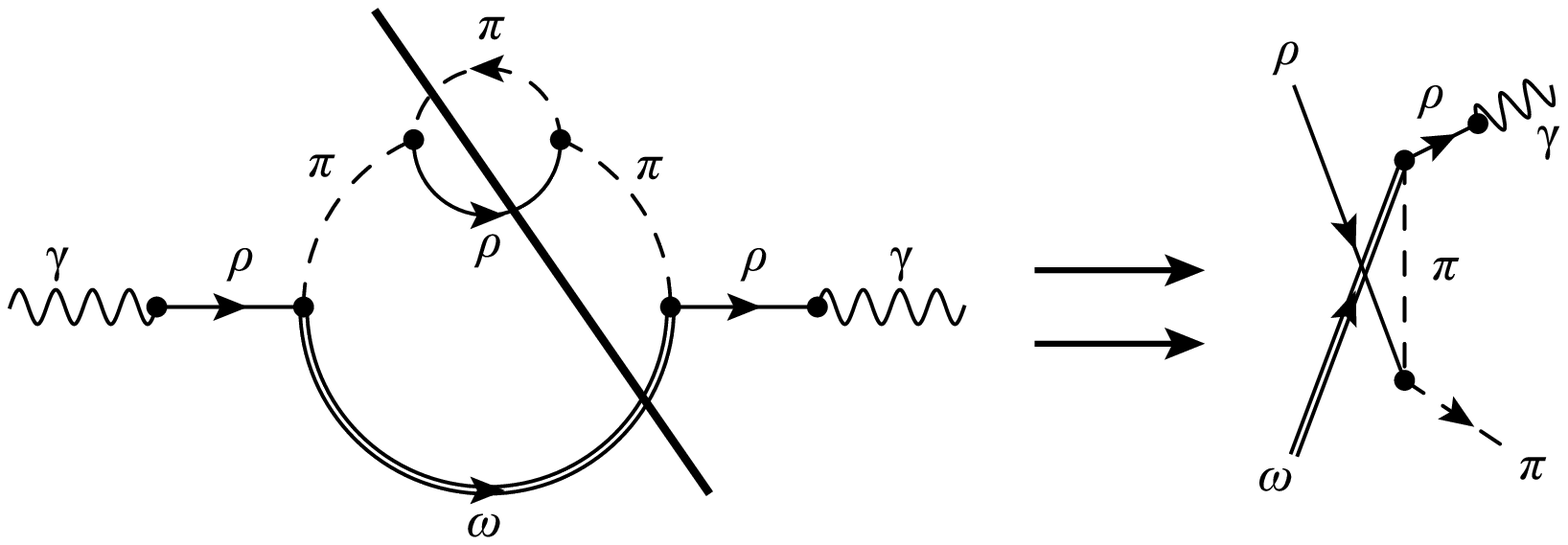}}
\subfigure[\, $\pi \omega \to \rho \gamma$ process]{
\includegraphics[scale=0.5]{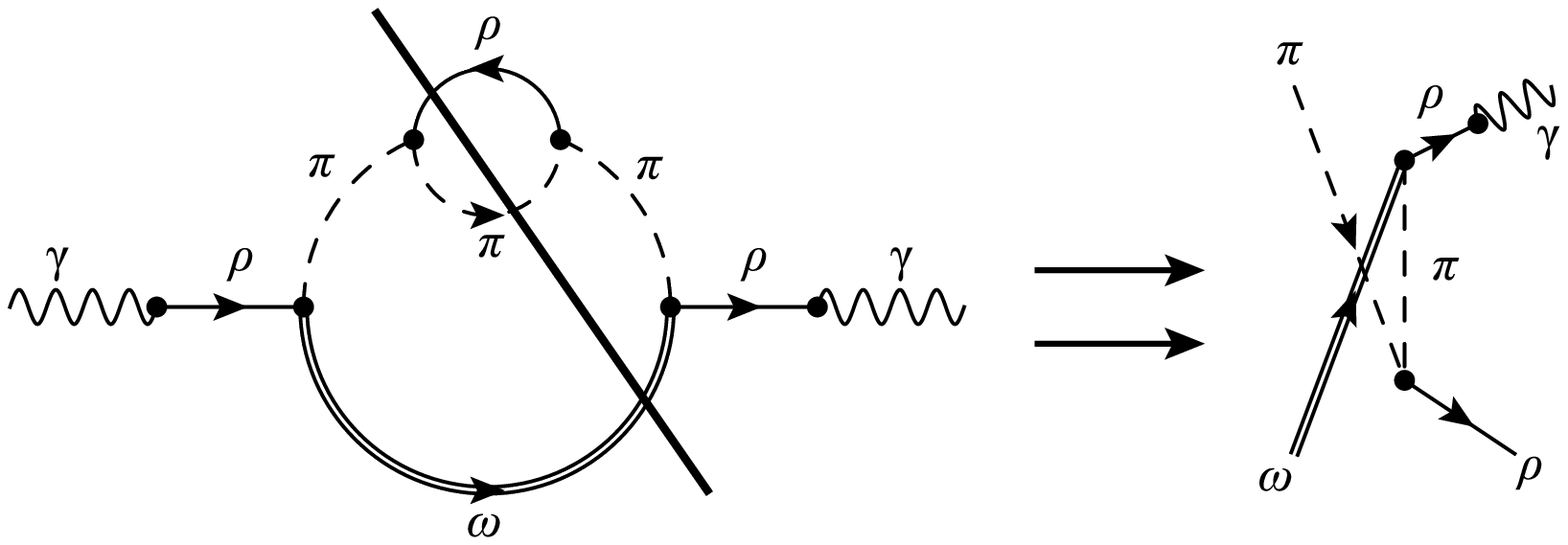}}
\caption{Cuts of the photon self-energy which generate imaginary
parts corresponding to the $u$-channel
diagrams of the $\rho \omega \to \gamma \pi$ and $\pi \omega \to \gamma \rho$ processes.}
\label{fig:rho-omega-cuts}
\end{figure}

For our analysis, the relevant self-energy diagrams are depicted
in Fig.~\ref{fig:rho-omega-cuts}, which encompass the $u$-channel diagrams
of the $\rho\omega \rightarrow \gamma\pi$ and
$\pi\omega\rightarrow\gamma\rho$ processes.
The latter is the diagram which involves the divergence and
double counting of the $\omega \to \pi^0 \gamma$
decay, while the former will be considered to benchmark
the equivalence between the two methods. For both cases, we can
separate the $u$-channel from the other channels and make
a meaningful comparison because each diagram is gauge-invariant
on its own due to the $\pi \rho \omega$ photo-emission vertex.

The two diagrams in Fig.~\ref{fig:rho-omega-cuts} have similar structures
differing only by the implementation of the $\pi$ self-energy in the inner loop.
Using standard Feynman rules within the Matsubara formalism~\cite{Kapusta:2006pm},
the (transverse part of the) $\rho$ self-energy encompassing both diagrams can
expressed as
\begin{equation}
\begin{split}
\label{eq:sigma-rho}
\Sigma_{\rho}^T(q,T)  = -\frac{1}{2} P_T^{\mu \mu'} \, g_{\pi\rho\omega}^2 \,
\int \frac{d^{3}p}{(2 \pi)^3} \, T \sum_{\omega_n}
v_{\pi \rho \omega}^{\mu\nu} v_{\pi\rho\omega}^{\mu'\nu'}
\\
\times D_{\omega}^{\nu \nu'}(\omega_n,\vec{p}\,)
D_{\pi}(q_0-\omega_n,\vec{q}-\vec{p}\,) \ ,
\end{split}
\end{equation}
where $v_{\pi \rho \omega}$ represents the $\pi \rho \omega$ vertex,
$\vec{p}$ is the three-momentum of the $\omega$ meson,
and $D_{\omega,\pi}$ are the propagators of the $\omega$ and $\pi$ mesons.
When evaluating $\Sigma_{\rho}^T$ at the photon point, the transverse projection
operator, $P_{T}^{\mu \nu}$, can be replaced by the (negative) metric tensor $-g^{\mu \nu}$,
as they are equivalent for a gauge invariant self-energy.

We will focus on calculating the imaginary part of the $\rho$ self-energy,
since the real part is small compared to the $\rho$ mass and thereby does not
significantly contribute to the rates.
First, each propagator is expressed in terms of a dispersion relation;
\begin{equation}
D(p_0,\vec{p}\,) = -\frac{1}{\pi} \int_{-\infty}^{\infty}
d\omega \frac{\mathrm{Im} \, D(\omega, \vec{p}\,)}{p_0 - \omega + i \epsilon} \ .
\end{equation}
The $\omega$ is considered a zero-width particle corresponding
to being an external particle in the KT calculation.
Using
\begin{equation}
{\rm Im} \, D_\omega (p_0,\vec{p}\,) = -\pi \delta\left(p_0^2
- \vec{p}^{\, 2} -m_\omega^2\right) \ ,
\end{equation}
 evaluating the Matsubara sum in Eq.~(\ref{eq:sigma-rho}), and
 taking the $\rho$ self-energy to the photon point, $q_0=|\vec{q}\,|$, yields
\begin{equation}
\begin{split}
\mathrm{Im} \, & \Sigma_{\rho} (q_0,\vec{q},T) = g_{\pi\rho\omega}^2 \int
\frac{d^3p}{(2\pi)^3} \frac{1}{2 E_{\omega}} \\
 & \times \bigg \{ (E_{\omega}q_0-\vec{p}\cdot\vec{q}\, )^2 \, \mathrm{Im} \,
 D_{\pi} \left(q_0-E_{\omega},\vec{q}-\vec{p}\,\right) \\
 & \, \, \, \, \, \, \, \big[1+ f^{\pi}(q_0-E_{\omega},T) +f^{\omega}(E_{\omega},T)   \big]
\\
& - (E_{\omega}q_0+\vec{p}\cdot\vec{q}\,)^2 \, \mathrm{Im} \, D_{\pi}
\left(E_{\omega}-q_0,\vec{q}-\vec{p}\,\right) \\
& \  \quad \, \big[f^{\pi}(E_{\omega}-q_0,T)-f^{\omega}(E_{\omega},T)  \big] \bigg \} \ ,
\end{split}
\label{eq:sigma-rho2}
\end{equation}
where $E_{\omega} = \sqrt{\vec{p}^{\, 2} + m_{\omega}^2}$.

The inner loop constitutes a pion self-energy which figures in the pion
propagators of Eq.~(\ref{eq:sigma-rho2}). For interactions with thermal
mesons, $m$, this pion self-energy has the form~\cite{Rapp:1999qu}
\begin{multline}
\Sigma_{\pi m}(k_0,\vec{k},T) = \int \frac{d^3p}{(2\pi)^3}
\frac{\mathcal{M}_{\pi m}(p,k_0,\vec{k}\,)}{2 E_m} \\
\times \Big\{f^m(E_m,T) - f^{\pi m}(E_m+k_0,T)\Big\}  \ ,
\label{eq:sigma-pi}
\end{multline}
where $\mathcal{M}_{\pi m}$ is the pertinent forward scattering amplitude.
Here we consider thermal mesons $m= \pi, \rho$, with either $\pi \pi$ scattering
through an $s$-channel $\rho$ resonance, or $\pi\rho$ scattering through an
$s$-channel $\pi$ resonance.
As indicated by the arrows in the diagrams of Fig.~\ref{fig:rho-omega-cuts},
taking the imaginary part of each $\rho$ self-energy yields the
corresponding $u$-channel Born diagram.
The $\pi$ self-energy $\Sigma_{\pi\pi}$ generates the
$\rho\omega\rightarrow\gamma\pi$
process, and the $\pi$ self-energy $\Sigma_{\pi\rho}$ generates
the $\pi\omega\rightarrow\gamma\rho$ process.

In the following two subsections we will analyze these two processes
in more detail. In Sec.~\ref{subsec:rhoomega}, we use the $\rho \omega \to \gamma \pi$
process as a benchmark to establish the equivalence of the photo-emission rates
calculated using TFT to 2-loop order with KT using Born amplitudes.  In
Sec.~\ref{subsec:piomega}, we then use the TFT calculation to identify a selection
criterion that allows us to eliminate double-counting with the $\omega$
radiative decay.

\subsection{$\rho \omega \to \gamma \pi$ $u$-channel}
\label{subsec:rhoomega}

\begin{figure}[!t]
\centering
\subfigure[\, Unitarity cut]{
\includegraphics[scale=0.5]{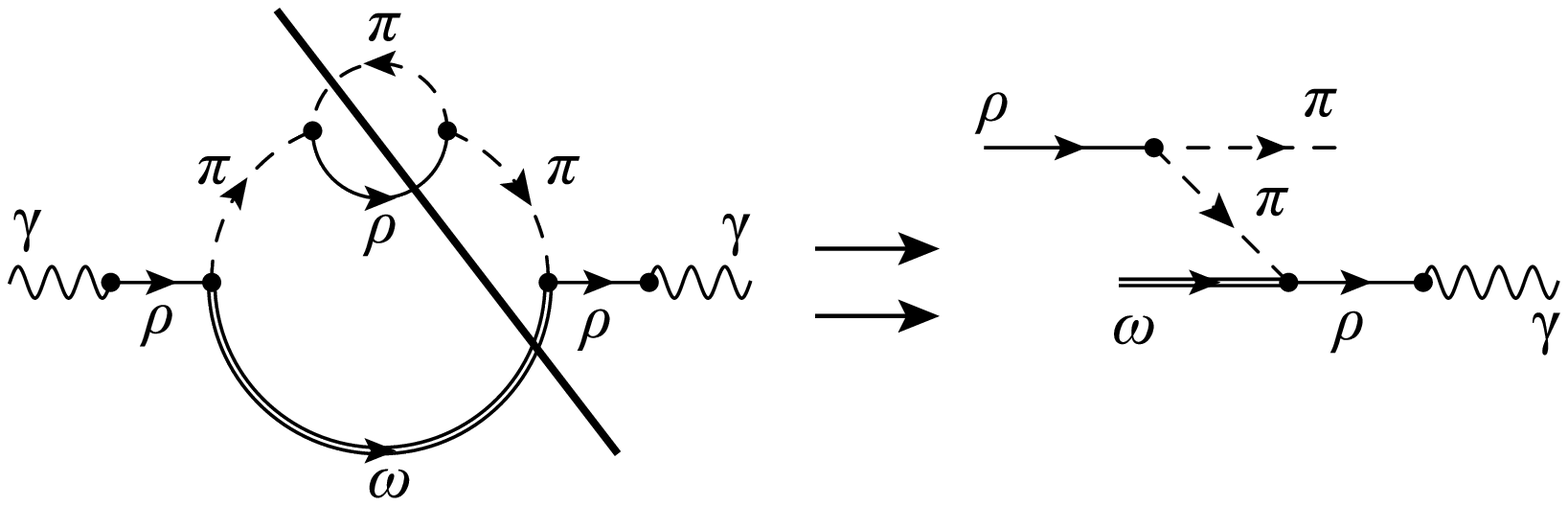}}
\subfigure[\, Landau cut]{
\includegraphics[scale=0.5]{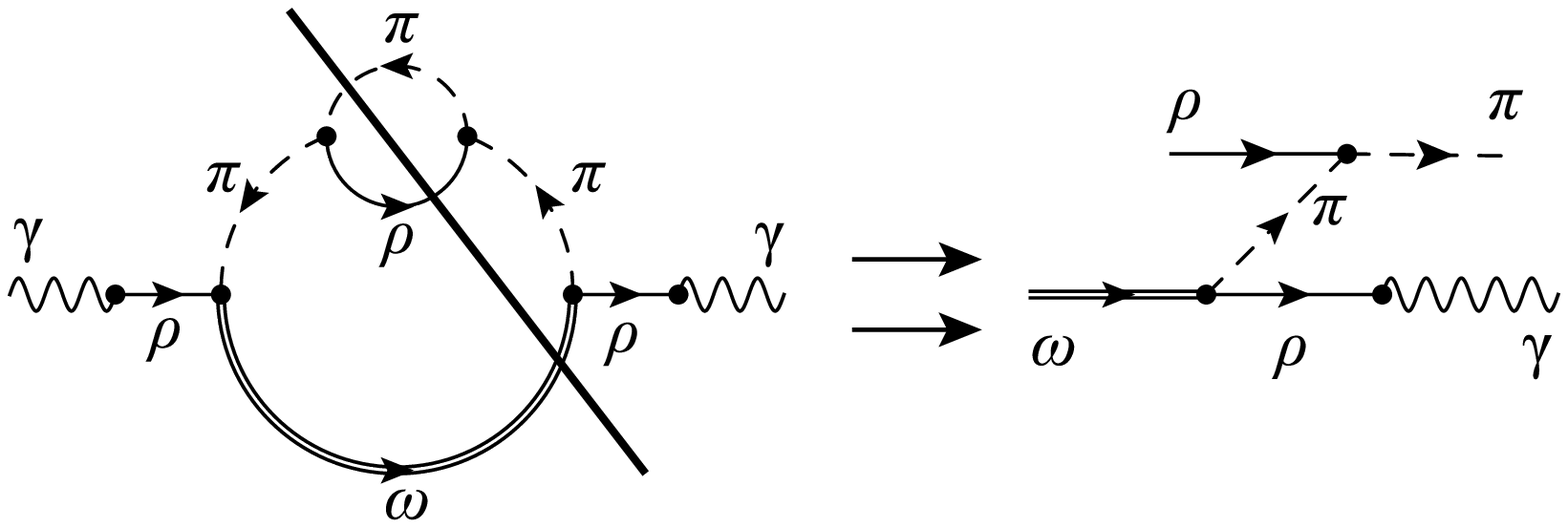}}
\caption{Two cuts of the photon self-energy which
give rise to imaginary parts
corresponding to the $u$-channel diagram of the
$\rho \omega \to \gamma \pi$ process.}
\label{fig:rho-cuts}
\end{figure}

While the $u$-channel diagram of the $\rho \omega \to \gamma \pi$
process is structurally similar to that of $\pi \omega \to \gamma \rho$
process, the key difference is that in the former the exchanged $\pi$ cannot go
on shell. This enables straightforward calculations for this process using
both KT and TFT.
The pertinent imaginary part of the $\rho$ self-energy has two major
contributions indicated by the two terms in braces in Eq.~(\ref{eq:sigma-rho2})
and schematically shown in Fig.~\ref{fig:rho-cuts}.  These two contributions,
commonly referred to as the Landau and unitarity cuts (representing
$\rho\pi\to\omega$ scattering and $\rho\to\pi\omega$ decay processes for
the left-hand-side vertex), can be interpreted based on the direction of energy
flow of the virtual pion. For the right-hand-side vertex, the unitarity cut is
associated with a pion energy flow oriented into the
$\pi \rho \omega$ vertex, as shown in Fig.~\ref{fig:rho-cuts}(a),
and is associated with $E_{\omega} < q_0$. This cut
corresponds to first term of Eq.~(\ref{eq:sigma-rho2}).
The Landau cut is associated with a pion energy flow oriented out of the
$\pi \rho \omega$ vertex, shown in Fig.~\ref{fig:rho-cuts}(b), and corresponds to
the second term of Eq.~(\ref{eq:sigma-rho2}), or $E_{\omega} > q_0$.

\begin{figure}[!t]
\centering
\includegraphics[width=.45\textwidth]{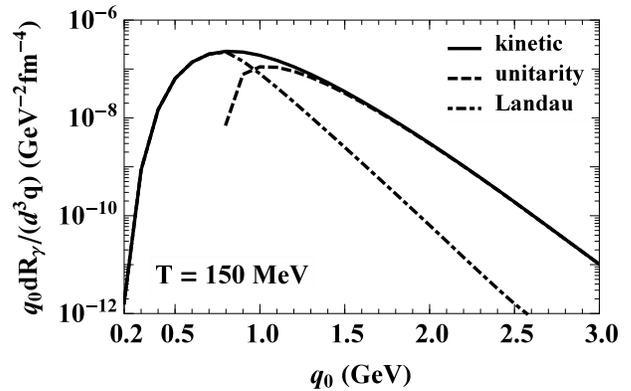}
\caption{Results from photo-emission calculation of
$\rho \omega \to \gamma \pi$ via the $u$-channel diagram at $T=150$\,MeV.
The solid line is the result using KT over the full kinematic range of the
exchanged pion, the dashed line is from TFT via the unitarity cut of
Fig.~\ref{fig:rho-cuts}, and the dot-dashed is from the Landau cut of
Fig.~\ref{fig:rho-cuts}. The sum of the unitarity and Landau cuts
is plotted but cannot be seen as it coincides with the solid curve.}
\label{fig:test-case}
\end{figure}

We have calculated the pertinent imaginary parts of the $\rho$ self-energy
given by Fig.~\ref{fig:rho-cuts} and inserted them into
Eq.~(\ref{eq:rate})\footnote{As in the KT calculations, we use $C_{\rho}$
as the EM $\rho\gamma$ coupling in Eq.~(\ref{eq:rate}) instead of the VMD
value of $e/g$.} to obtain the corresponding photo-emission
rates; the results for a temperature of $T=150$\,MeV are shown in
Fig.~\ref{fig:test-case} and compared to the KT calculation (both without
form factors). We find very good agreement between the sum of the Landau
and unitarity cuts in TFT and the KT Born amplitude calculation, thus
confirming the equivalence of the two methods.
In addition, we have verified that the Landau (unitarity) cut contribution in
the TFT calculation can be directly mapped to a calculation in KT where
the phase space integral is restricted to the energy of the exchanged pion flowing
out of (in to) the $\pi \rho \omega$ vertex ($E_{\omega} >$ or $< q_0$,
respectively).
The identification of the mapping between Landau and unitarity cuts and KT
for this process ($\rho \omega \to \gamma \pi$) is facilitated by the fact
that the $u$-channel Born diagram does not develop any on-shell singularities
(contrary to the $\pi \omega \to \gamma \rho$ process discussed below).

Figure~\ref{fig:rho-cuts}(b) shows that the Landau cut gives rise to a Born
diagram featuring a $\omega\to\pi\gamma$ radiative decay topology.
However, the emitted $\pi$ is necessarily spacelike as it is absorbed by an
on-shell $\rho$ turning into an on-shell pion. Likewise, in the Born diagram
generated from the unitarity cut an on-shell $\omega$ absorbs a $\pi$ turning
into a massless photon, requiring a spacelike $\pi$.
Similar considerations for the unitarity and Landau cuts will be used in
the following section  to separate out on-shell $\omega$ decays in the
$\pi \omega \to \gamma \rho$ process.

The TFT calculation affords us with yet another benefit, namely to evaluate the
effects of a finite width on the $\rho$ mesons which appear as external particles
in the Born calculations within kinetic theory. This effect has been studied
previously within the latter framework and found to be negligible~\cite{Alam:2001ar}.
Within TFT we can render this check more rigorous by including a finite width for the
$\rho$ meson in the integral for the $\pi\rho$ loop in Fig.~\ref{fig:rho-cuts}(a).
We found variations of order 10\% or less of the photoemission rates, corroborating
the results of Ref.~\cite{Alam:2001ar}.

\subsection{$\pi \omega \to \gamma \rho$ $u$-channel}
\label{subsec:piomega}
We now utilize TFT to calculate the photo-emission rate given by the
$\rho$ self-energy in Fig.~\ref{fig:rho-omega-cuts}(b) where the inner
loop is a pion self-energy induced by an interaction with a thermal $\rho$
meson forming a $\pi$, i.e., $\Sigma_{\pi\rho}$ from Eq.~(\ref{eq:sigma-pi}).
Unlike its KT counterpart, this calculation is well defined, without any
divergences associated with, e.g., a pion pole.

\begin{figure}[!t]
\centering
\subfigure[\, Unitarity cut]{
\includegraphics[scale=0.5]{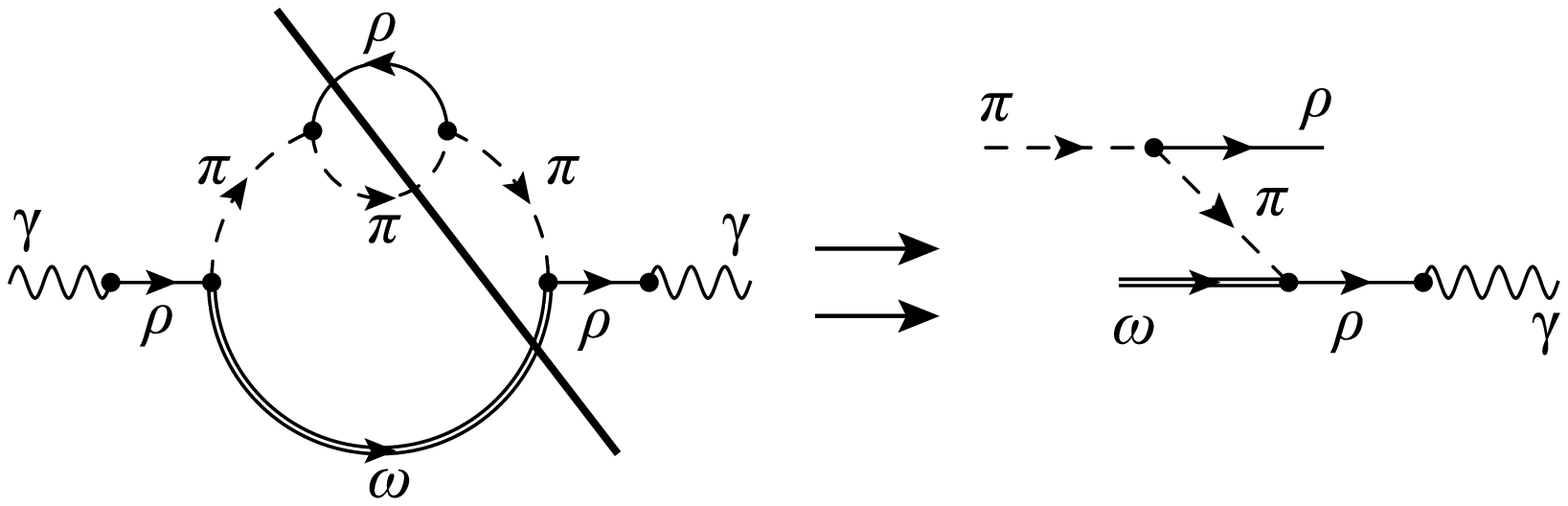}}
\subfigure[\, Landau cut]{
\includegraphics[scale=0.5]{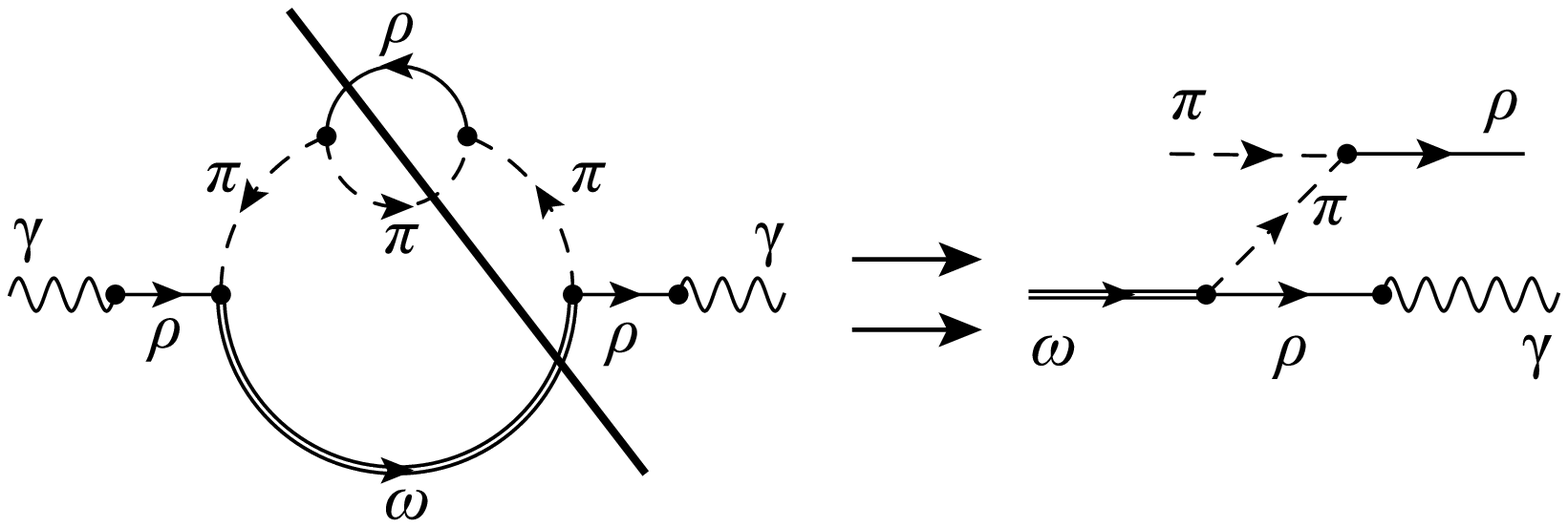}}
\caption{Two cuts of the photon self-energy which give rise to imaginary
parts corresponding to the $u$-channel diagram of the
$\pi \omega \to \gamma \rho$ process. }
\label{fig:cuts}
\end{figure}

Again, as illustrated in Fig.~\ref{fig:cuts}, the $\rho$ self-energy can
be separated into unitarity and Landau cuts. However, in this process the
exchanged $\pi$ in the Landau cut can go on-shell. When this occurs the Landau
cut corresponds to the radiative $\omega$ decay of $\omega \to \pi^0 \gamma$.
This contribution has already been included in previous
calculations of thermal photo-emission rates~\cite{Rapp:1999qu}, and does not
constitute a new hadronic source from the $\pi\rho\omega$ system.

We now use the distinction between the Landau and unitarity cuts provided by TFT
as a selection criterion for what to include in our rates. We will drop the
contribution from the Landau cut altogether, corresponding to the exchange of
timelike pions, to preclude any double-counting with the radiative $\omega$ decay.
The remaining contribution from the unitarity cut calculated with TFT
is found to agree with KT when the integration over phase space is restricted
to pion energies such that $q_0>E_\omega$.
In principle, this is a conservative choice since it excludes not only all
timelike pions but also spacelike ones with positive energy. In practice, this
difference appears to be negligible.

\section{Thermal Photon Rates from $\pi\rho\omega$ Interactions}
\label{sec:results}
\begin{figure}[t!]
\centering
\includegraphics[width=.45\textwidth]{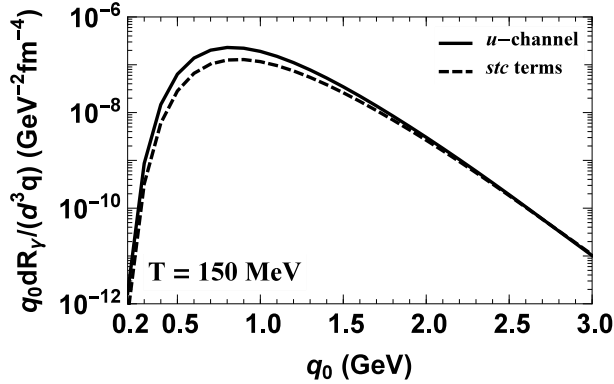}
\caption{Comparison of the contributions to the
$\rho \omega \to \gamma \pi$ process from the $u$-channel diagram
(solid line) and the combined $stc$ terms (dashed line); no
form factors included.}
\label{fig:rho-omega-u-stc}
\end{figure}

\begin{figure}[!t]
\centering
\includegraphics[width=.45\textwidth]{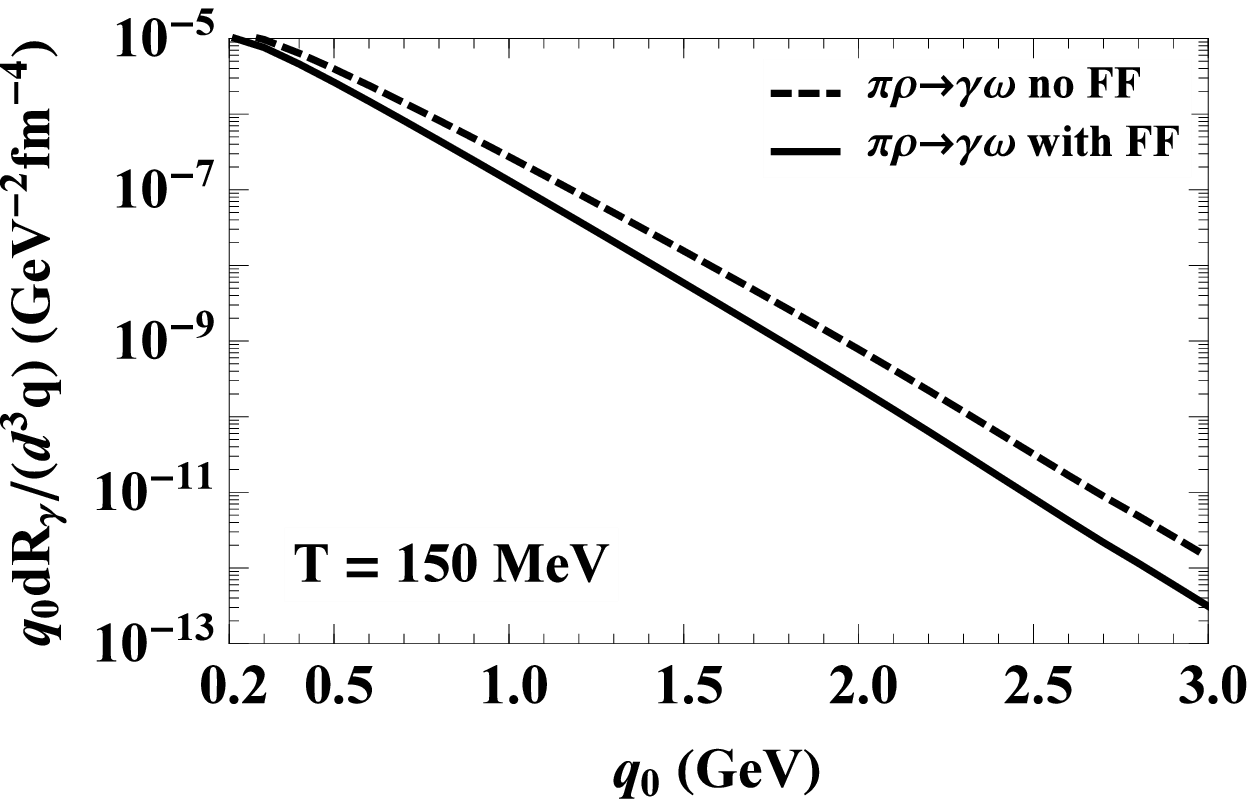}
\includegraphics[width=.45\textwidth]{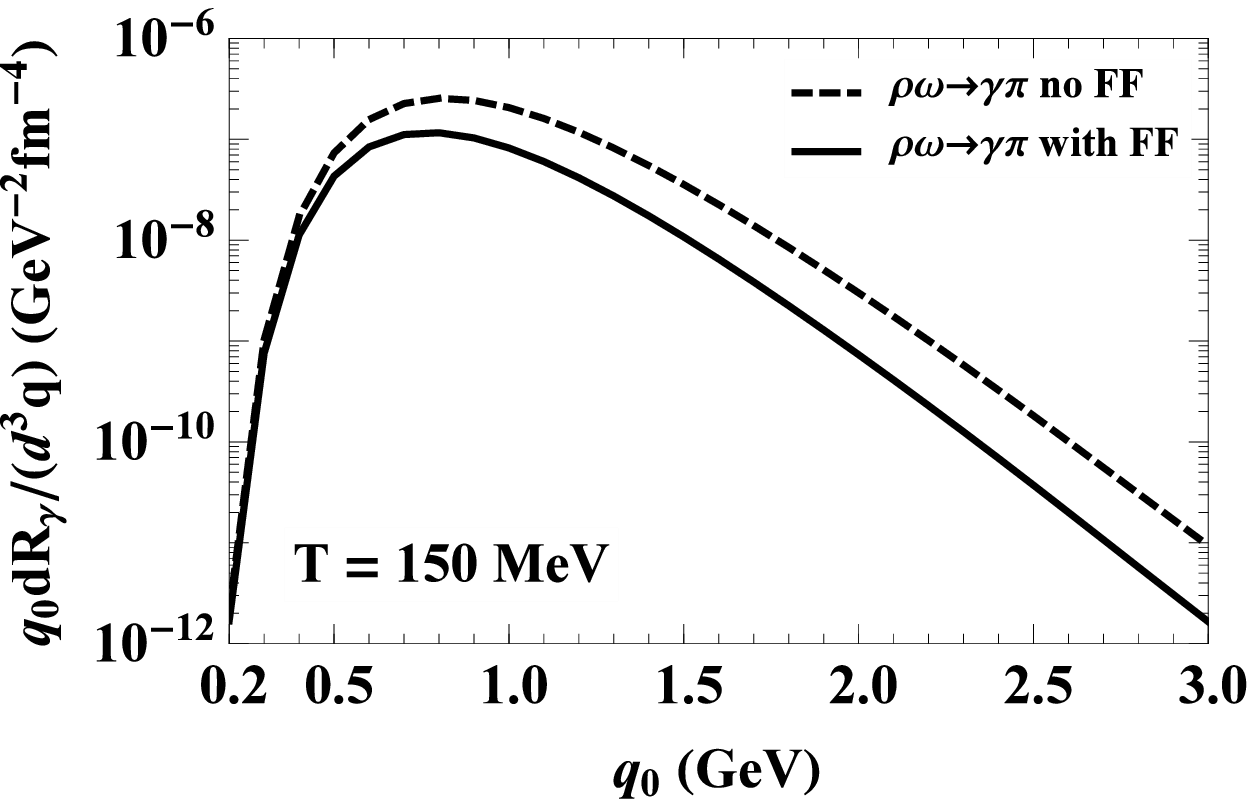}
\includegraphics[width=.45\textwidth]{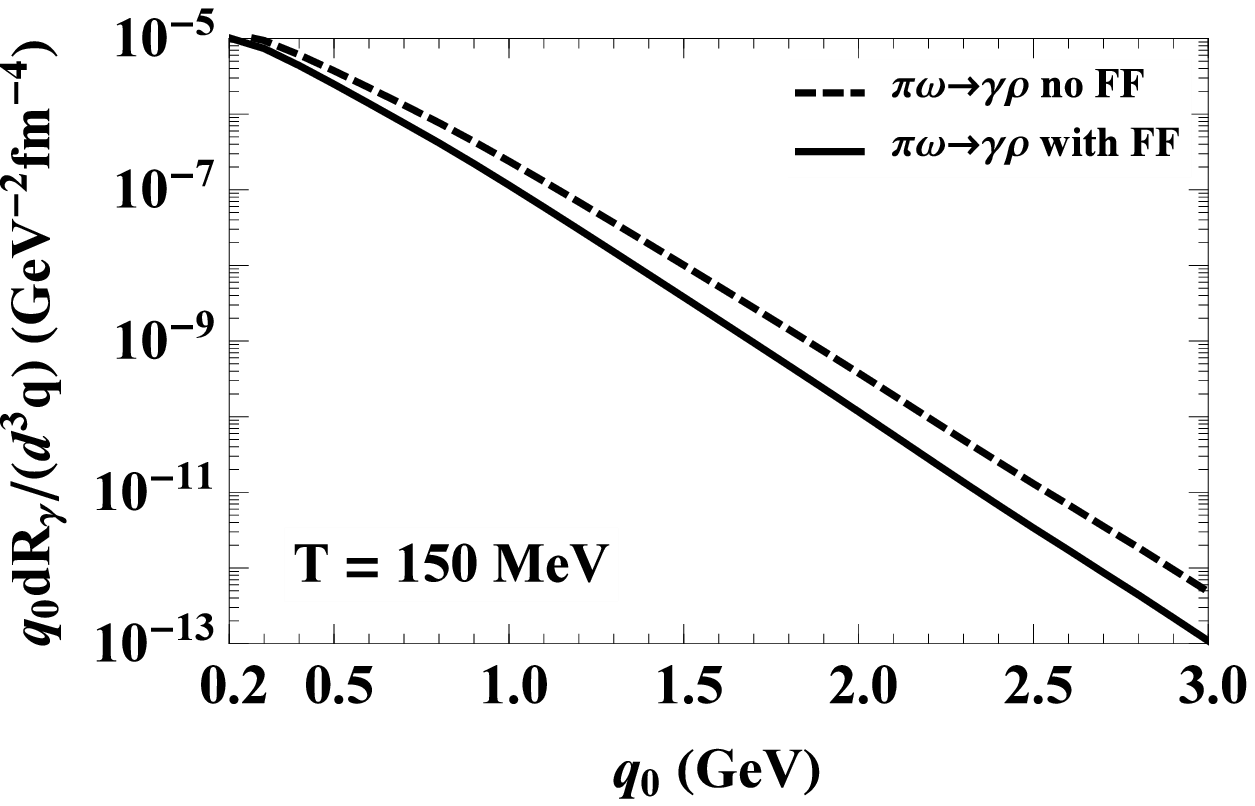}
\caption{Impact of hadronic form factors on the photo-emission rates at $T=150$ MeV
for $\pi \rho \to \gamma \omega$ (upper panel), $\rho \omega \to \gamma \pi$
(middle panel), and $\pi \omega \to \gamma \rho$ (lower panel) processes. The rates
with form factor (solid lines) are compared to the ones without form factor (dashed
lines).}
\label{fig:ff}
\end{figure}

Before coming to our final results, let us  briefly elaborate on
the concrete implementation of form factors for the processes at hand.
The $\pi \omega \to \gamma \rho$ process is dominated by the $t$- and
$u$-channel pion exchanges.
Since the $t$- and $u$-channel form factors have the same structure, their
factorized average form factor is identical and is thus applied as overall
form factor to all diagrams in this process.
In the $\pi \rho \to \gamma \omega$ system,
the pion $t$-channel exchange is expected to prevail over the $u$-channel, which
is suppressed by the $\rho$ mass in the propagator. Therefore the pertinent factorized
average $t$-channel form factor is employed.
Before implementation of form factors, the $\rho \omega \to \gamma \pi$ process
has two approximately equal contributions: the $u$-channel and the combined
contact, $s$-, and $t$-channel terms (``$stc$'' for brevity),
cf.~Fig.~\ref{fig:rho-omega-u-stc}. Note that both $u$ and the $stc$ diagrams
are individually gauge invariant which affords us the possibility to treat
these two contributions separately. As discussed in Sec.~\ref{sec:photons},
the suppression generated by the averaged factorized form factors is driven
by the mass of the exchanged particle. The $u$-channel diagram involves an
exchanged $\pi$ whose associated form factor generates a suppression of
up to a factor 4.5 at $q_0 = 3.0$\,GeV. On the other hand, the $stc$ term is
dominated by $t$-channel $\rho$ exchange at high energies, whose associated
average form factor generates a suppression of up to a factor 30 at
$q_0 = 3.0$~GeV. Clearly, the choice of either from factor would not be
an accurate procedure. However, due to the separate gauge invariance of the $u$
and $stc$ terms,  we can apply an average $\pi$ exchange form factor ($F_\pi$)
to the $u$-channel, an average $\rho$ exchange form factor ($F_\rho$) to
the $stc$ term,
and a combination of the two form factors to the interference term
(which is also gauge invariant), schematically given by
 \begin{equation}
 \begin{split}
 |\mathcal{M}_{\mathrm{FF}}|^2 =& F_{\pi}^4 |\mathcal{M}_u|^2
 + F_{\rho}^4 |\mathcal{M}_{stc}|^2 \\ &+ F_{\pi}^2 F_{\rho}^2
 \left( \mathcal{M}_u  \mathcal{M}_{stc}^{*} +  \mathcal{M}_{stc}
 \mathcal{M}_{u}^{*} \right) \ .
 \end{split}
\end{equation}
The net effect of this implementation for the $\rho \omega \to \gamma \pi$
process is that the total rate is suppressed by a somewhat larger magnitude
than the other two processes, but still less suppressed than if we had
used an overall $t$-channel $\rho$-exchange form factor.
The quantitative effects of the form factors are illustrated in
Fig.~\ref{fig:ff} for each of the three processes at a temperature of
$T=150$\,MeV.

One may wonder how variations in the form factor cutoff affect the
emission rates. The cutoff parameter of $\Lambda_{\pi\rho\omega}=1$\,GeV
(as used throughout this work) has been fixed together with the $\pi\rho\omega$
coupling constant in Ref.~\cite{Rapp:1999qu}, to reproduce the radiative
and hadronic $\omega$ decays. However, a somewhat smaller cutoff, say,
$\Lambda_{\pi\rho\omega}=0.8$\,GeV, still gives agreement with those data
once the coupling constant is increased accordingly. We have verified that
using the latter set of parameters leads to an insignificant change of
our photon emission rates over the relevant range of photon energies of up
to $\approx$5\,GeV.

\begin{figure}[t!]
\centering
\includegraphics[width=.45\textwidth]{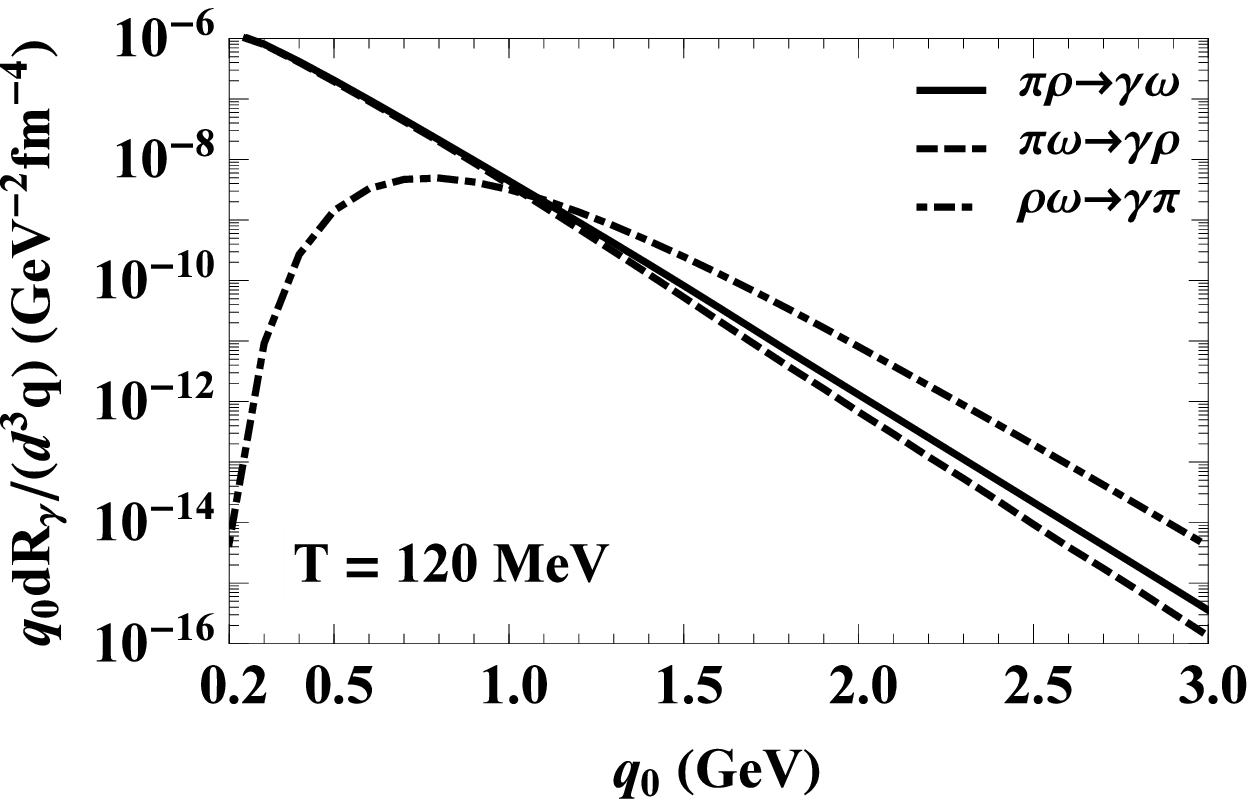}
\includegraphics[width=.45\textwidth]{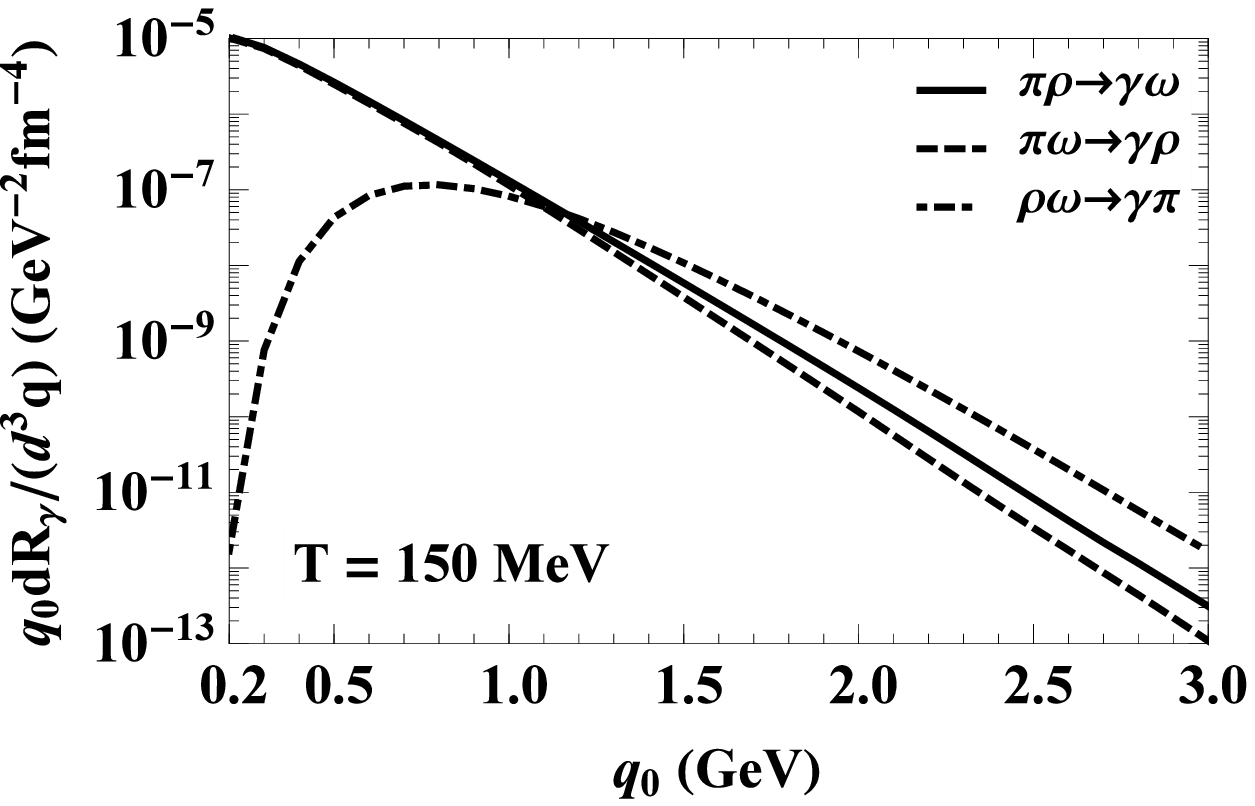}
\includegraphics[width=.45\textwidth]{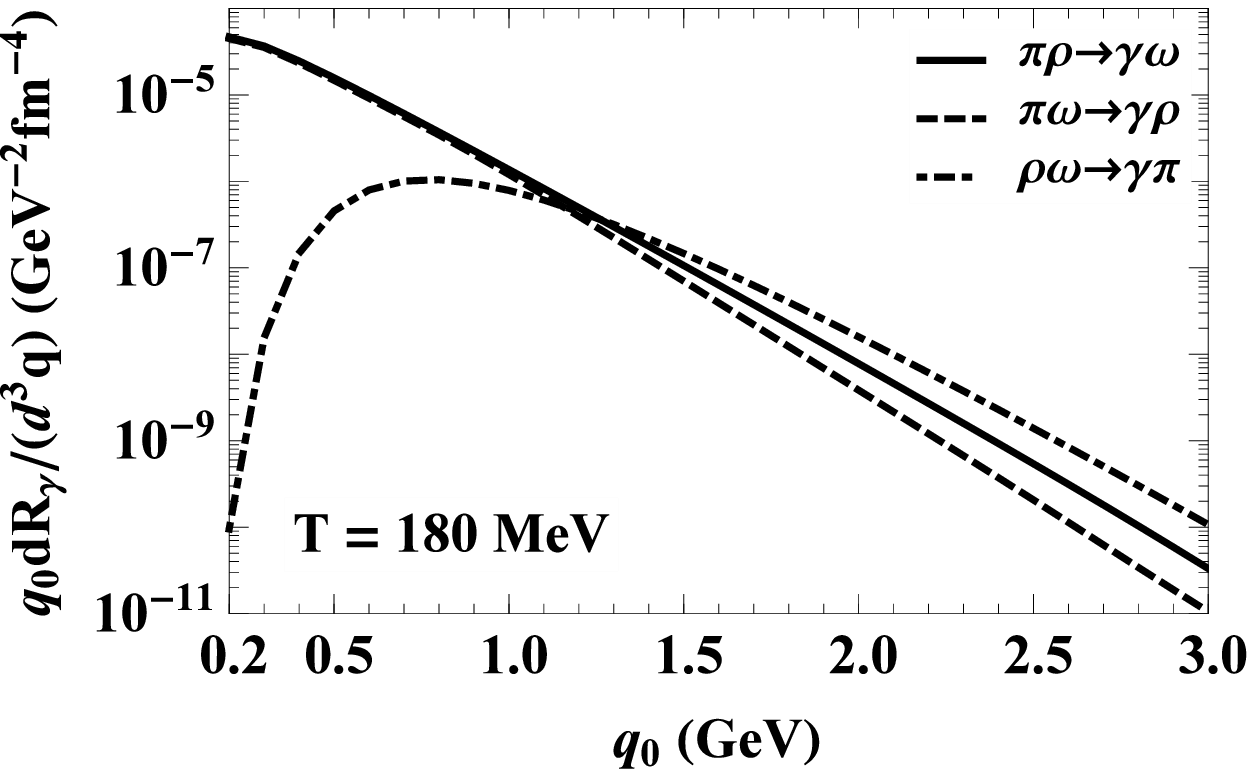}
\caption{Thermal photo-emission rates from the $\pi \rho \omega$ system for
$\pi \rho \to \gamma \omega$ (solid lines),  $\pi \omega \to \gamma \rho$
(dashed lines), and $\rho \omega \to \gamma \pi$ (dot-dashed lines) for
temperatures $T=120$, 150 and 180\,MeV (upper, middle and lower panel,
respectively). Form factor effects are included.}
\label{fig:total-sep}
\end{figure}

Our final photo-emission rates for the three processes from the $\pi \rho \omega$
system (including form factors) are summarized in Fig.~\ref{fig:total-sep} for three
different temperatures of 120, 150, and 180 MeV.
In the phenomenologically important regime of photon
energies around $\approx$1\,GeV all three channels are of comparable
magnitude. The $\rho \omega \to \gamma \pi$ channel falls off significantly
below that, but becomes the dominant photon source at energies above
1.2-1.5\,GeV for all temperatures. The relative spectral strengths of the
three channels are rather stable with temperature; only the
$\pi \rho \to \gamma \omega$ rate increases slightly relative to the other two
at higher energies.  We provide parametrizations of the final rates in
Appendix~\ref{app:parametrizations}.
While the applicability of the
present methods of thermal photon calculations is questionable at temperatures
close to the pseudo-critical transition temperature, we include parametrizations
up to $T=180$\,MeV for usage in a variety of thermal evolution models
which may have the phase transition implemented at different temperatures.

\begin{figure}[t!]
\centering
\includegraphics[width=.45\textwidth]{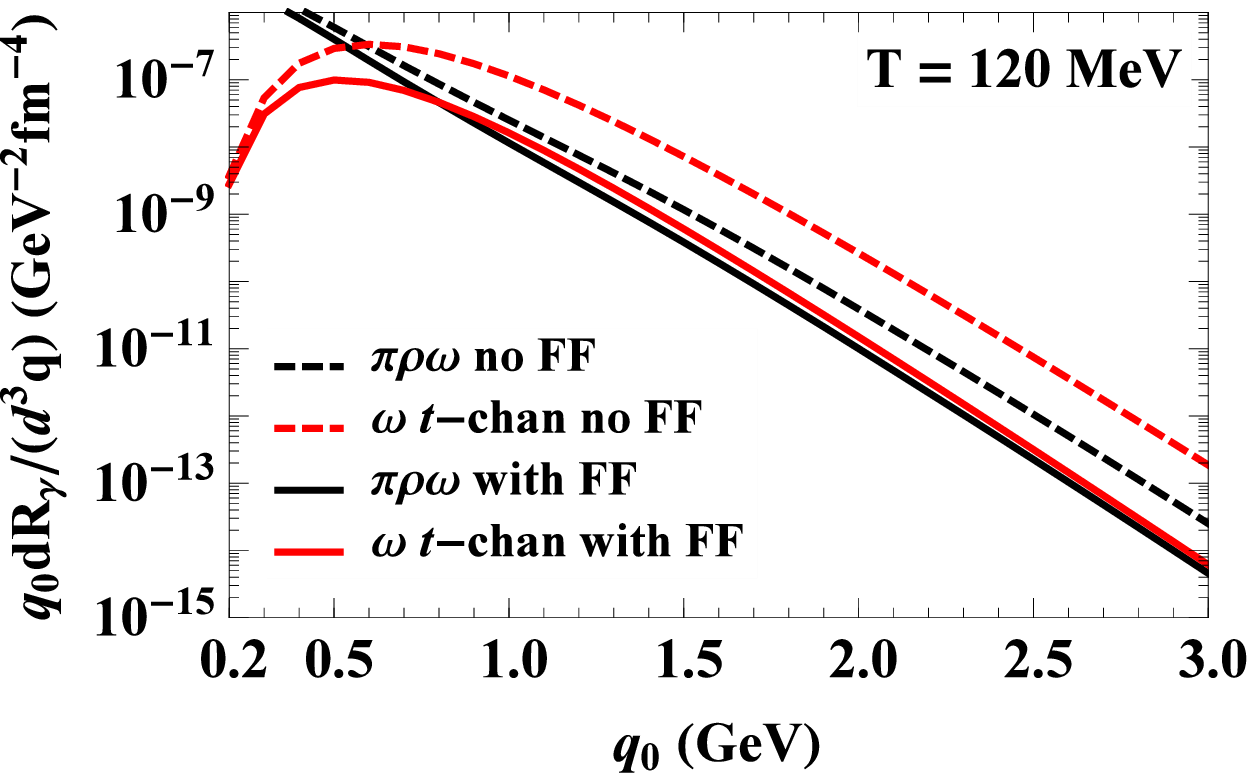}
\includegraphics[width=.45\textwidth]{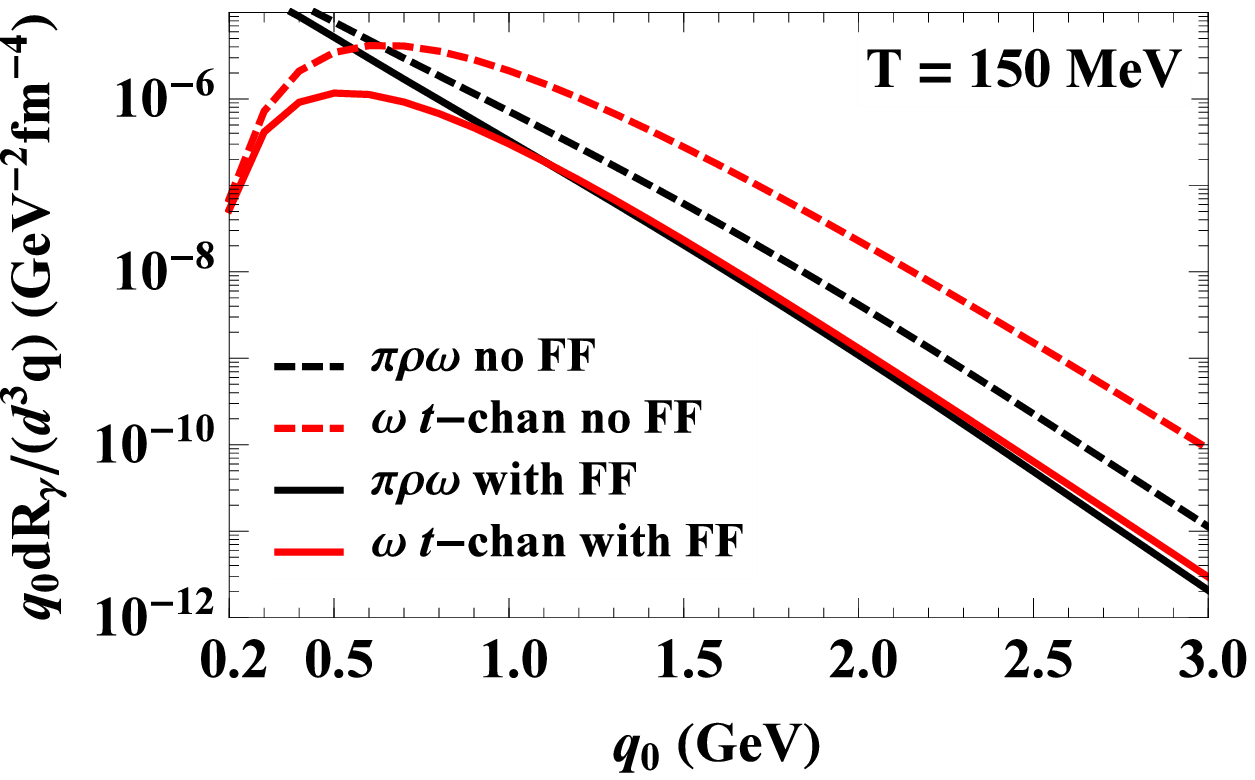}
\includegraphics[width=.45\textwidth]{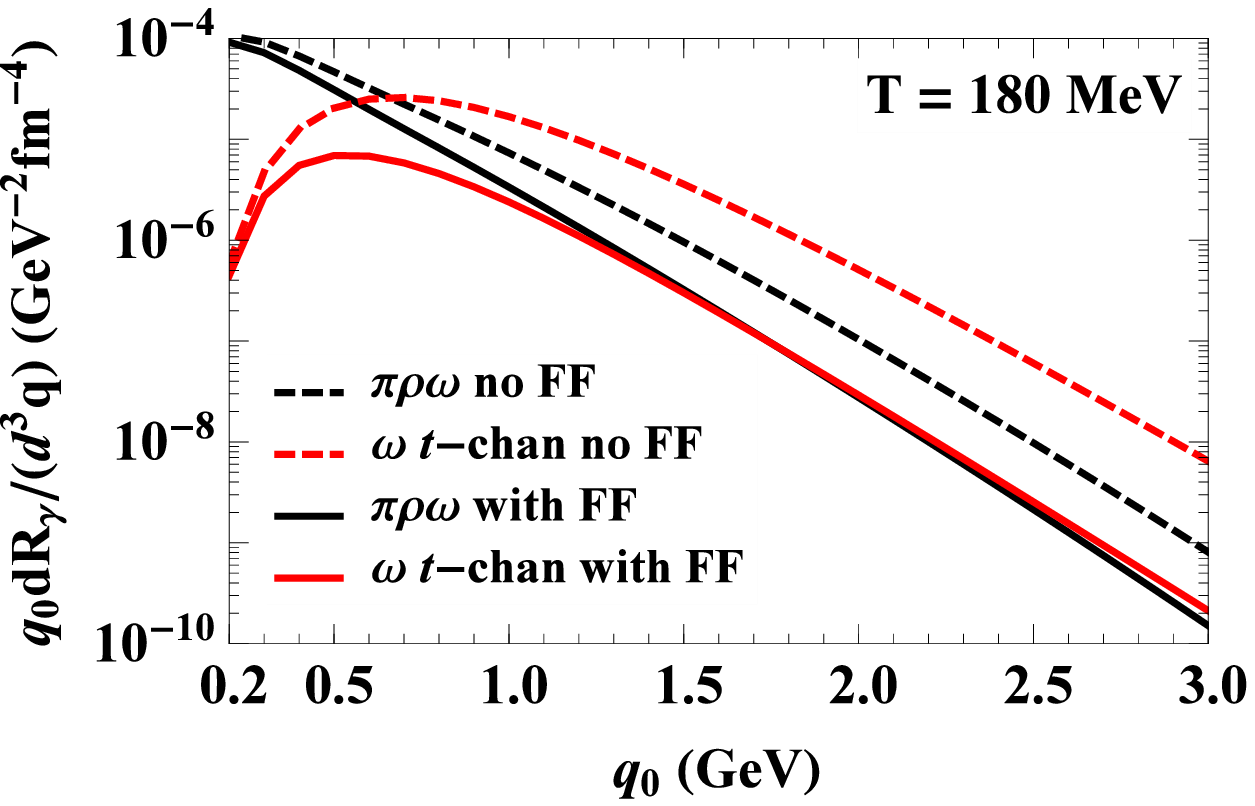}
\caption{Total rates from the $\pi \rho \omega$ system as calculated
in the present work (black lines) versus the $\omega$ $t$-channel rate
(red line) for temperatures $T=120$, 150 and 180\,MeV
(upper, middle and lower panel, respectively). Dashed lines are without
form factor; solid lines are with form factor.}
\label{fig:total}
\end{figure}

\begin{figure}[b]
\centering
\includegraphics[width=.45\textwidth]{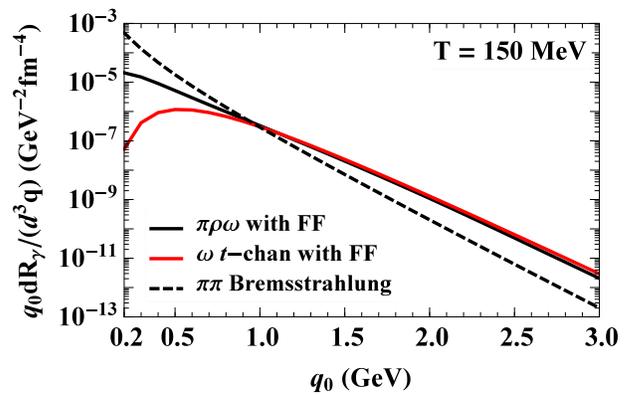}
\caption{Total rates at $T=150$\,MeV from the $\pi \rho \omega$ system (solid
black line) compared to the $\pi \pi$ Bremsstrahlung rate (dashed black
line)~\cite{Liu:2007zzw,Heffernan:2014mla} and the $\pi\rho\to\gamma\pi$
$\omega$ $t$-channel rate (red line).}
\label{fig:total-brem}
\end{figure}

At last, we compare the total rate from the newly calculated processes with
existing literature, specifically, the $\omega$ $t$-channel exchange in
$\pi \rho \to \gamma \pi$ as calculated in Ref.~\cite{Turbide:2003si}
(Fig.~\ref{fig:total}) and
 $\pi\pi$ Bremsstrahlung from Refs.~\cite{Liu:2007zzw,Heffernan:2014mla}
(Fig.~\ref{fig:total-brem}). The former process involves the same vertices as
considered in the present work but with the $\omega$ as an exchange particle
rather than an external one; the pertinent
rate was found to be a significant source of thermal photons for energies
$q_0\ge 1.5$\,GeV relative to other known sources in hot and dense hadronic
matter~\cite{Turbide:2003si,Heffernan:2014mla}.
Prior to the inclusion of form factors, the sum of the newly calculated rates
is smaller than the $\omega$ $t$-channel exchange in the $\pi \rho \to \gamma \pi$
process by a factor of ca.~4-8 for $q_0> 1$\,GeV. In the realistic case with
form factors, however, the two rates are rather close in the phenomenologically
most relevant range of $q_0<2$\,GeV. In practice, in URHICs at RHIC and LHC
energies, this approximately translates into lab-momenta of $q_t<4$\,GeV due
to blue-shift effects of the exploding fireball~\cite{Rapp:2014qha}.
In Refs.~\cite{Liu:2007zzw,Heffernan:2014mla}, $\pi\pi$ Bremsstrahlung was
found to be appreciable for photon energies of less than 1\,GeV, even exceeding
the contribution from in-medium $\rho$ mesons with baryonic
sources~\cite{Urban:1999im,Rapp:1999us}
at the lowest photon energies.
Fig.~\ref{fig:total-brem} shows that the rates from the $\pi \rho \omega$
system are comparable to the Bremsstrahlung rates for photon energies between
0.5 and 1\, GeV, which suggests that the contribution of this new photon source
is significant relative to existing thermal photon rate calculations.

In addition, below the chemical freezeout temperature of $T_{\rm ch}\simeq160$
in URHICs, effective meson chemical potentials build up, in particular
for pions, which further augment two of the three newly calculated rates.
The $\pi \omega \to \gamma \rho$ and $\rho \omega \to \gamma \pi$ processes
will pick up pion fugacity factors $z_\pi=\exp(\mu_\pi/T)$ to the
$4^{\mathrm{th}}$ and $5^{\mathrm{th}}$ powers, respectively, compared to
the $3^{\mathrm{rd}}$ power for $\pi \rho \to \gamma \pi$ processes.
With this enhancement from pion fugacities, the thermal photon emission
processes calculated could provide a non-negligible contribution to the
direct-photon spectra in URHICs~\cite{vanHees:2014ida}.

In our current study, we have focused on the $\pi\rho\omega$ system
due to its known large coupling constant and the relatively small masses
of these hadrons. Decay contributions of higher resonances as intermediate
particles in hadron-$\rho$ scattering processes have been considered in
Ref.~\cite{Rapp:1999qu} and found to give subleading contributions at the
photon point (cf.~Fig.~1 in that reference). The question remains whether
the scattering of higher-mass states can give significant contributions.
As an example, let us consider the $a_1$ as an external particle. This is
the next higher-mass particle with a known large coupling to $\pi\rho$ and
$\pi\gamma$, and closely resembles the $\pi\rho\omega$ system in the pertinent
photon-emitting processes.  Inspection of Fig.~1 of Ref.~\cite{Rapp:1999qu}
reveals that the $a_1$ contribution to the $\rho$ self-energy is down by
about an order of magnitude
compared to the $\omega$ at the photon point. This factor can be readily
understood by realizing that the $\pi\rho\omega$ coupling is a factor 2
larger than the $\pi\rho a_1$ coupling, and that the thermal $a_1$
density is suppressed by more than a factor of 3 at temperatures of
$T=150$\,MeV.

\section{Conclusions}
\label{sec:conclusion}
In this work, we have calculated the photo-emission rates from the
tree-level scattering processes $\pi \rho \to \gamma \omega$,
$\pi \omega \to \gamma \rho$, and $\rho \omega \to \gamma \pi$
using relativistic kinetic theory. Complementary calculations were
performed using thermal field theory
for the $u$-channel diagrams of two of the processes. This allowed
us to (a) explicitly establish consistency between the two methods
(as well as exert quality control of the results) and (b) identify a
criterion by which to exclude singular contributions from the
exchange of on-shell pions in $\pi \omega \to \gamma \rho$ and thus
avoid double-counting with the $\omega$ radiative decay. After the
inclusion of hadronic vertex form factors, which suppress the rates
at high energies, our total rate resulting from all three processes
turns out to be comparable to the $\omega$ $t$-channel rates in the
$\pi \rho \to \gamma \pi$ process~\cite{Turbide:2003si} and to
$\pi\pi$ Bremsstrahlung~\cite{Liu:2007zzw}. Our results thus
provide an enhancement of existing rate calculations in the direction
of what has been conjectured in Ref.~\cite{vanHees:2014ida};
we anticipate that thermal emission
from $\pi\rho\omega$ interactions will give a sizable contribution
to direct-photon spectra in heavy-ion collisions. The precise extent
to which the new thermal photon
sources may help to reduce discrepancies with experimental spectra
and elliptic flow will require their implementation into evolution
models for the fireball in heavy-ion collisions.
Work in this direction is in progress.

\acknowledgments
This work is supported by the US National Science Foundation
under grant no. PHY-1306359.

\appendix
\onecolumngrid
\section{Born Matrix Elements}
\label{app:amplitudes}
In this appendix, we write out the matrix elements corresponding to each
Feynman diagram in the $\pi \rho \omega$ system, explicitly indicating
isospin indices and labeling each particle's four-momentum.  They are
derived by applying standard Feynman rules to the diagrams shown in
Fig.~\ref{fig:diagrams} and using the interaction Lagrangians of
Eqs.~(\ref{eq:interaction1})-(\ref{eq:EM-coupling}).
\\

$\bullet$ $\pi^a(p_1) \, \rho^b(p_2) \to \gamma(q) \, \omega(p_3)$ process with isospin indices $\pi^a \rho^b$:
\begin{figure}[h!]
\centering
\includegraphics[scale=0.5]{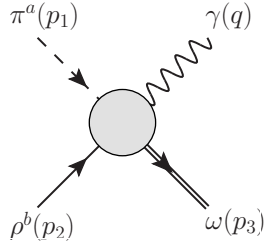}
\caption{Diagram showing four-momenta labels and isospin indices for the process
$\pi \rho \to \gamma \omega$. }
\label{fig:rho-blob}
\end{figure}
\begin{eqnarray}
\mathcal{M}_s &=& -\frac{g_{\pi\rho\omega} g_{\rho} C_{\rho}}{s-m_{\pi}^2} (2p_1-p_2)_{\delta} q^{\mu} q^{\alpha}
\, \epsilon_{\mu \gamma \alpha \beta} \, \varepsilon^{\delta}(p_2) \varepsilon^{*\gamma}(q) \varepsilon^{*\beta}(p_3) \, \epsilon^{3ab} \\
\mathcal{M}_t &=& -\frac{g_{\pi\rho\omega} g_{\rho} C_{\rho}}{t-m_{\pi}^2} (2p_1-q)_{\gamma} p_2^{\mu} q^{\alpha}
\, \epsilon_{\mu \delta \alpha \beta} \, \varepsilon^{\delta}(p_2) \varepsilon^{*\gamma}(q) \varepsilon^{*\beta}(p_3) \, \epsilon^{3ab} \\
\mathcal{M}_u &=& -\frac{g_{\pi\rho\omega} g_{\rho} C_{\rho}}{t-m_{\rho}^2} (p_2-q)^{\mu} q^{\alpha}
\left( -g^{\nu\lambda} + \frac{(p_2-q)^{\nu}(p_2-q)^{\lambda}}{m_{\rho}^2} \right) \\
 && \left[ -g_{\delta \gamma}(p_2+q)_{\lambda} - g_{\gamma \lambda}(p_2-2q)^{\delta} + g_{\delta \lambda}(2p_2-q)_{\gamma} \right]
\, \epsilon_{\mu \nu \alpha \beta} \, \varepsilon^{\delta}(p_2) \varepsilon^{*\gamma}(q) \varepsilon^{*\beta}(p_3) \, \epsilon^{3ab} \nonumber \\
\mathcal{M}_c &=& -g_{\pi\rho\omega} g_{\rho} C_{\rho} q^{\alpha} \, \epsilon_{\gamma \delta \alpha \beta} \, \varepsilon^{\delta}(p_2) \varepsilon^{*\gamma}(q)
\varepsilon^{*\beta}(p_3) \, \epsilon^{3ab}
\end{eqnarray}
\\

$\bullet$ $\pi^a(p_1) \, \omega(p_2) \to \gamma(q) \, \rho^b(p_3)$ process with isospin indices $\pi^a \rho^b$:
\begin{figure}[h!]
\centering
\includegraphics[scale=0.5]{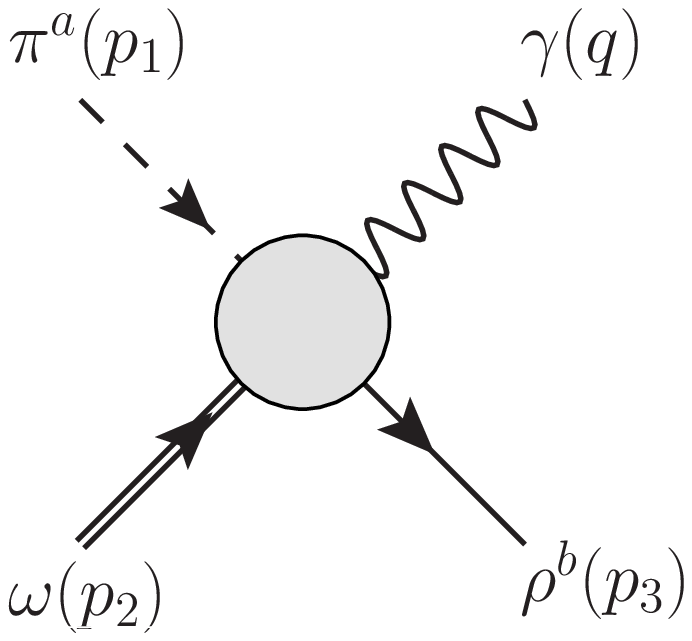}
\caption{Diagram showing four-momenta labels and isospin indices for the
process $\pi \omega \to \gamma \rho$. }
\label{fig:omega-blob}
\end{figure}
\begin{eqnarray}
\mathcal{M}_s &=& -\frac{g_{\pi\rho\omega} g_{\rho} C_{\rho}}{s-m_{\rho}^2} p_1^{\mu} p_2^{\alpha}
\left( -g^{\nu\lambda} + \frac{(q+p_3)^{\nu}(q+p_3)^{\lambda}}{m_{\rho}^2} \right) \\
&&\left[ g_{\lambda \delta}(q+2p_3)_{\gamma} - g_{\lambda \gamma}(2q+p_3)_{\delta} + g_{\gamma \delta}(q-p_3)_{\lambda} \right]
\, \epsilon_{\mu \nu \alpha \beta} \, \varepsilon^{\beta}(p_2) \varepsilon^{*\gamma}(q) \varepsilon^{*\delta}(p_3) \epsilon^{3ab} \nonumber   \\
\mathcal{M}_t &=& \frac{g_{\pi\rho\omega} g_{\rho} C_{\rho}}{t-m_{\pi}^2} (p_1-q)^{\mu} p_2^{\alpha} (2p_1-q)_{\gamma}
\, \epsilon_{\mu \delta \alpha \beta} \, \varepsilon^{\beta}(p_2) \varepsilon^{*\gamma}(q) \varepsilon^{*\delta}(p_3) \, \epsilon^{3ab}  \\
\mathcal{M}_u &=& \frac{g_{\pi\rho\omega} g_{\rho} C_{\rho}}{u-m_{\pi}^2} (p_2-q)^{\mu} p_2^{\alpha} (p_1-p_2+q)_{\delta}
\, \epsilon_{\mu \gamma \alpha \beta} \, \varepsilon^{\beta}(p_2) \varepsilon^{*\gamma}(q) \varepsilon^{*\delta}(p_3) \, \epsilon^{3ab}  \\
\mathcal{M}_c &=& g_{\pi\rho\omega} g_{\rho} C_{\rho} p_2^{\alpha} \, \epsilon_{\delta \gamma \alpha \beta} \,
\varepsilon^{\beta}(p_2) \varepsilon^{*\gamma}(q) \varepsilon^{*\delta}(p_3) \, \epsilon^{3ab}
\end{eqnarray}
\\

$\bullet$ $\rho^a(p_1) \, \omega(p_2) \to \gamma(q)\, \pi^b(p_3)$ process
with isospin indices $\rho^a \pi^b$:
\begin{figure}[h!]
\centering
\includegraphics[scale=0.5]{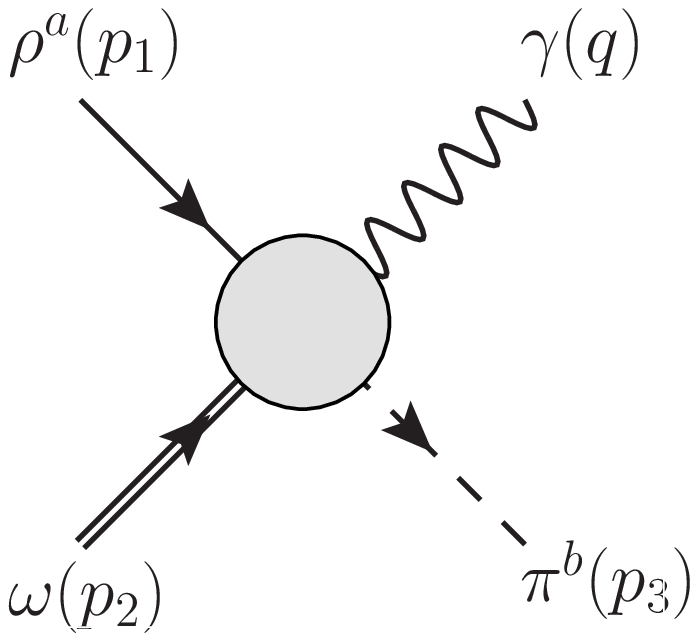}
\caption{Diagram showing four-momenta labels and isospin indices for
the process $\rho \omega \to \gamma \pi$. }
\label{fig:rho-omega-blob}
\end{figure}
\begin{eqnarray}
\mathcal{M}_s &=& \frac{g_{\pi\rho\omega} g_{\rho} C_{\rho}}{s-m_{\pi}^2} p_1^{\mu} p_2^{\alpha}
(q+2p_3)_{\gamma} \, \epsilon_{\mu \delta \alpha \beta} \, \varepsilon^{\delta}(p_1)
\varepsilon^{\beta}(p_2) \varepsilon^{*\gamma}(q) \, \epsilon^{3ab} \\
\mathcal{M}_t &=& -\frac{g_{\pi\rho\omega} g_{\rho} C_{\rho}}{t-m_{\rho}^2} (p_1-q)^{\mu}  p_2^{\alpha}
\left( -g^{\nu \lambda} + \frac{(p_1-q)^{\nu}(p_1-q)^{\lambda}}{m_{\rho}^2} \right) \\
 &&\left[ -g_{\delta \gamma}(p_1+q)_{\lambda} - g_{\gamma \lambda}(p_1-2q)_{\delta} + g_{\delta \lambda}(2p_1-q)_{\gamma}  \right]
 \, \epsilon_{\mu \nu \alpha \beta} \, \varepsilon^{\delta}(p_1) \varepsilon^{\beta}(p_2) \varepsilon^{*\gamma}(q) \, \epsilon^{3ab} \nonumber \\
\mathcal{M}_u &=& \frac{g_{\pi\rho\omega} g_{\rho} C_{\rho}}{u-m_{\pi}^2} p_3^{\mu} p_2^{\alpha} (p_2-q+p_3)_{\delta}
\, \epsilon_{\mu\gamma\alpha\beta} \, \varepsilon^{\delta}(p_1) \varepsilon^{\beta}(p_2) \varepsilon^{*\gamma}(q) \, \epsilon^{3ab} \\
\mathcal{M}_c  &=& g_{\pi\rho\omega} g_{\rho} C_{\rho} p_2^{\alpha} \, \epsilon_{\delta \gamma \alpha \beta} \, \varepsilon^{\delta}(p_1)
\varepsilon^{\beta}(p_2) \varepsilon^{*\gamma}(q) \, \epsilon^{3ab}
\end{eqnarray}

\section{Parametrizations}
\label{app:parametrizations}
In this appendix, we present parametrizations of the photo-emission rates
for each process in the $\pi \rho \omega$ system, along with plots of
comparisons of parametrizations to calculated rates.  We have verified the
accuracy of the parametrizations to within 10\% for temperature and photon
energy ranges of
$100 \, \mathrm{MeV} \le T \le 180 \, \mathrm{MeV}$ and
$0.2 \, \mathrm{GeV} \le q_0 \le 5.0 \, \mathrm{GeV}$, except for the lowest
photon energies of the $\rho \omega \to \gamma \pi$ process, whose
overall contribution in that photon energy range is negligible. Form factor effects
are included in all rate parametrizations.
\\

\scalebox{1.2}{$\bullet$ \underline{$\pi \rho \to \gamma \omega$}}
\begin{figure}[h!]
\centering
\includegraphics[width=.45\textwidth]{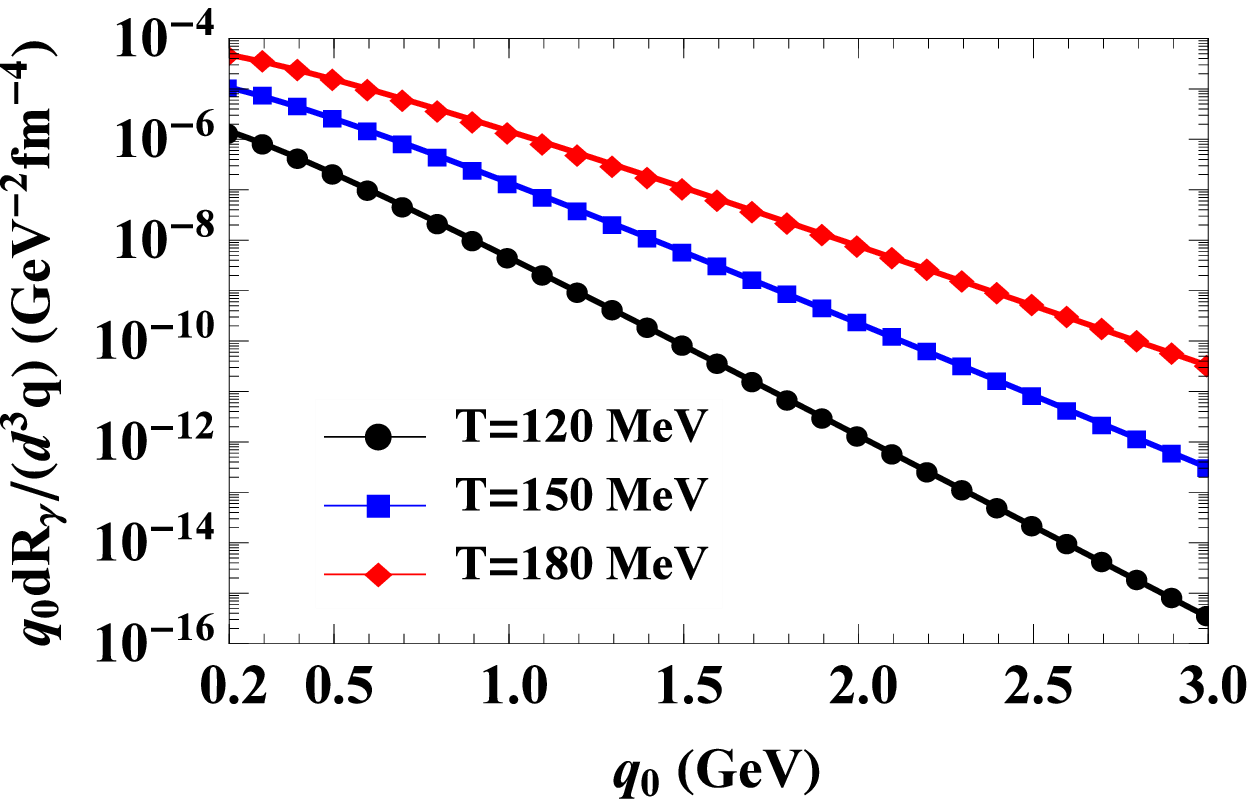}
\includegraphics[width=.435\textwidth]{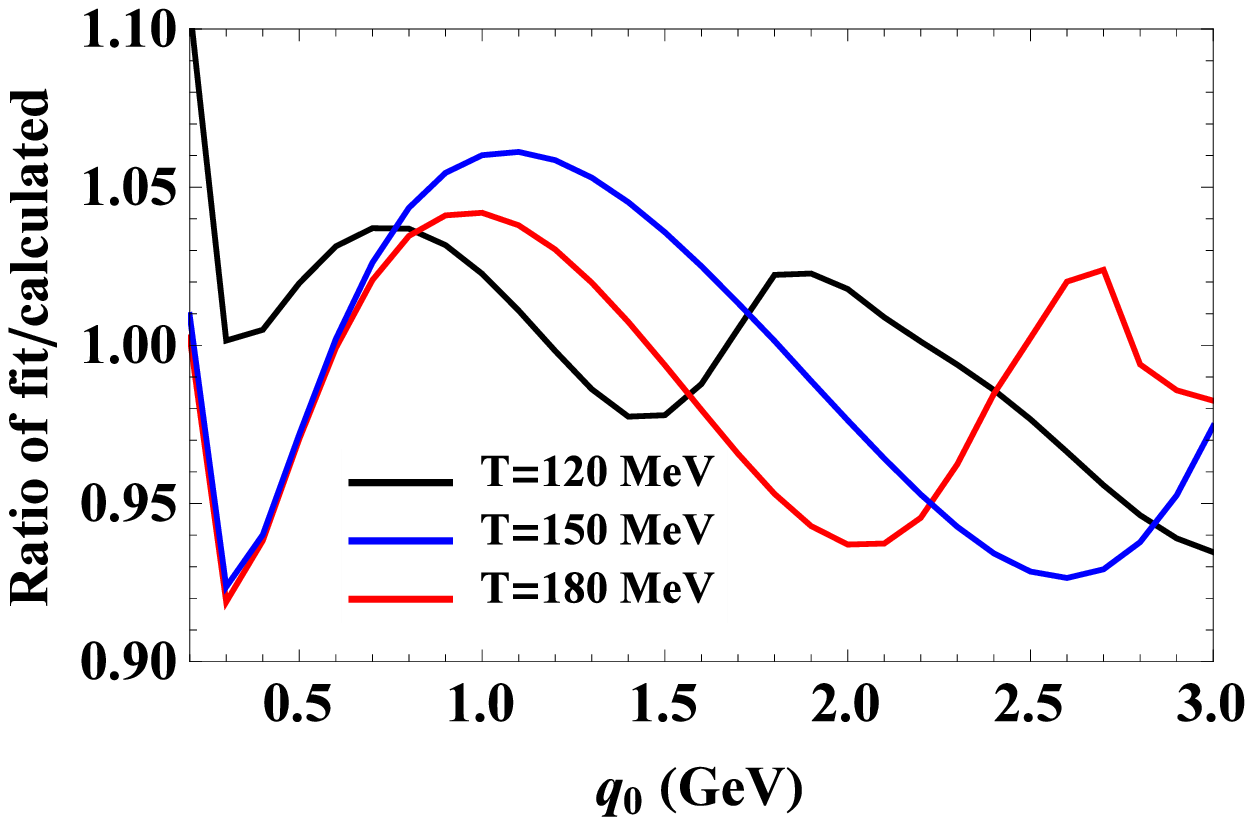}
\caption{Left panel: Calculated thermal photo-emission rates from the
$\pi \rho \to \gamma \omega$ process (symbols) compared to the parametrized
rates (lines). Right panel: Ratio of parametrized rates to
calculated rates.}
\end{figure}

\begin{eqnarray}
 q_0 \frac{dR_{\pi \rho \to \gamma \omega}}{d^3q} & = &  \exp
\big[ a_1q_0 + a_2 +a_3q_0^{a_4}
 + a_5(q_0 + a_6)^{a_7} \big]
\, \left[\mathrm{fm}^{-4} \, \mathrm{GeV}^{-2} \right] \ ,  \\
 a_1(T) & = &  -35.8991 + 460.425 \,T - 2592.04 \,T^2 + 5342.32\, T^3 \ , \nonumber \\
 a_2(T) & = &  -41.9725 + 601.952 \,T - 3587.8 \,T^2 + 7604.97 \,T^3 \ , \nonumber \\
 a_3(T) & = &  0.740436 - 16.7159 \,T + 133.526 \,T^2 - 347.589\, T^3 \ , \nonumber \\
 a_4(T) & = & 2.00611 - 3.79343 \,T + 29.3101 \,T^2 - 72.8725 \,T^3 \ , \\
 a_5(T) & = &  -8.33046 + 121.091 \,T - 801.676\, T^2 + 1712.16\, T^3 \ , \nonumber \\
 a_6(T) &  = &  17.9029 - 388.5 \,T + 2779.03 \,T^2 - 6448.4\, T^3 \ , \nonumber \\
 a_7(T) & = &  -15.622 + 340.651 \,T - 2483.18\, T^2 + 5870.61\, T^3 \ \nonumber .
\end{eqnarray}
\\

\scalebox{1.2}{$\bullet$ \underline{$\pi \omega \to \gamma \rho$}}
\begin{figure}[h!]
\centering
\includegraphics[width=.45\textwidth]{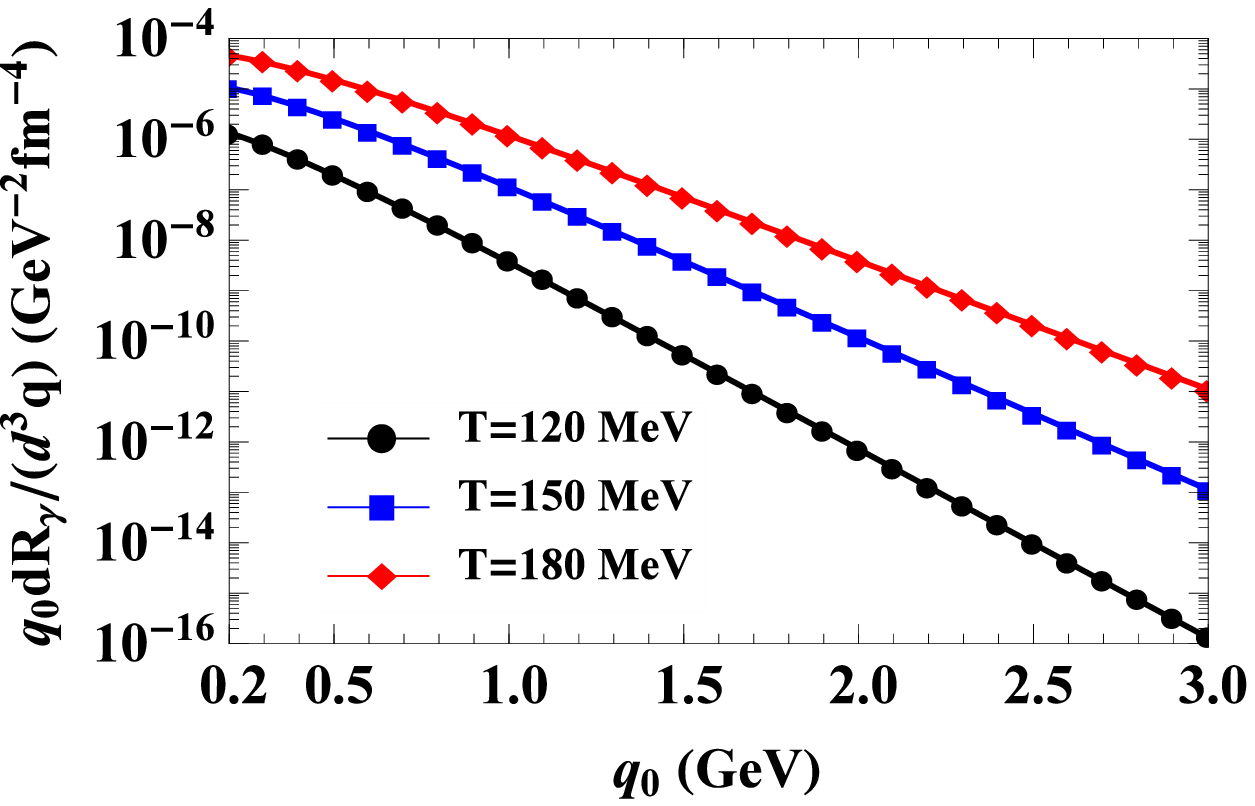}
\includegraphics[width=.435\textwidth]{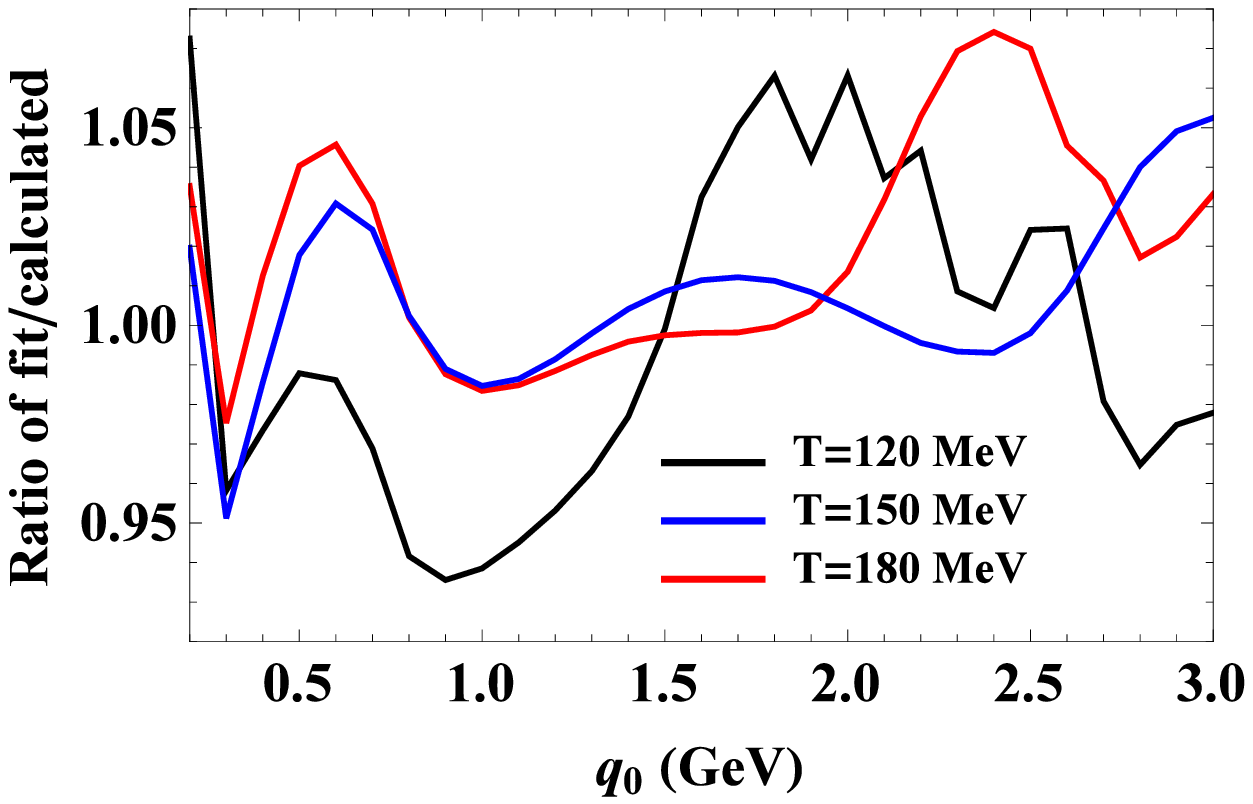}
\caption{Left panel: Calculated thermal photo-emission rates from the
$\pi \omega \to \gamma \rho$ process (symbols) compared to the parametrized
rates (lines). Right panel: Ratio of parametrized rates to
calculated rates.}
\end{figure}
\begin{eqnarray}
 q_0 \frac{dR_{\pi \omega \to \gamma \rho}}{d^3q} & = &  \exp
\big[ a_1q_0 + a_2 +a_3q_0^{a_4}
 + a_5(q_0 + a_6)^{a_7} \big]
\, \left[\mathrm{fm}^{-4} \, \mathrm{GeV}^{-2} \right] \ ,  \\
 a_1(T) & = &  -29.4663 + 291.356 \, T - 1301.27 \,T^2 + 2102.12\, T^3 \ , \nonumber \\
 a_2(T) & = & -45.081 + 688.929 \,T - 4150.15 \,T^2 + 8890.76 \,T^3 \ , \nonumber \\
 a_3(T) & = &  -0.260076 + 8.92875\, T - 60.868 \,T^2 + 136.57 \,T^3 \ , \nonumber \\
 a_4(T) & = & 2.2663 - 8.30596 \,T + 49.3342 \,T^2 - 90.8501 \,T^3 \ , \\
 a_5(T) & = &  10.2955 - 317.077 \,T + 2412.15 \,T^2 - 6020.9 \,T^3 \ , \nonumber \\
 a_6(T) &  = &  3.12251 - 47.5277 \,T + 222.61 \,T^2 - 241.9 \,T^3 \ , \nonumber \\
 a_7(T) & = &  -3.39045 + 56.5927 \,T - 336.97\, T^2 + 622.756 \,T^3 \ \nonumber .
\end{eqnarray}
\\

\scalebox{1.2}{$\bullet$ \underline{$\rho \omega \to \gamma \pi$}}
\begin{figure}[h!]
\centering
\includegraphics[width=.45\textwidth]{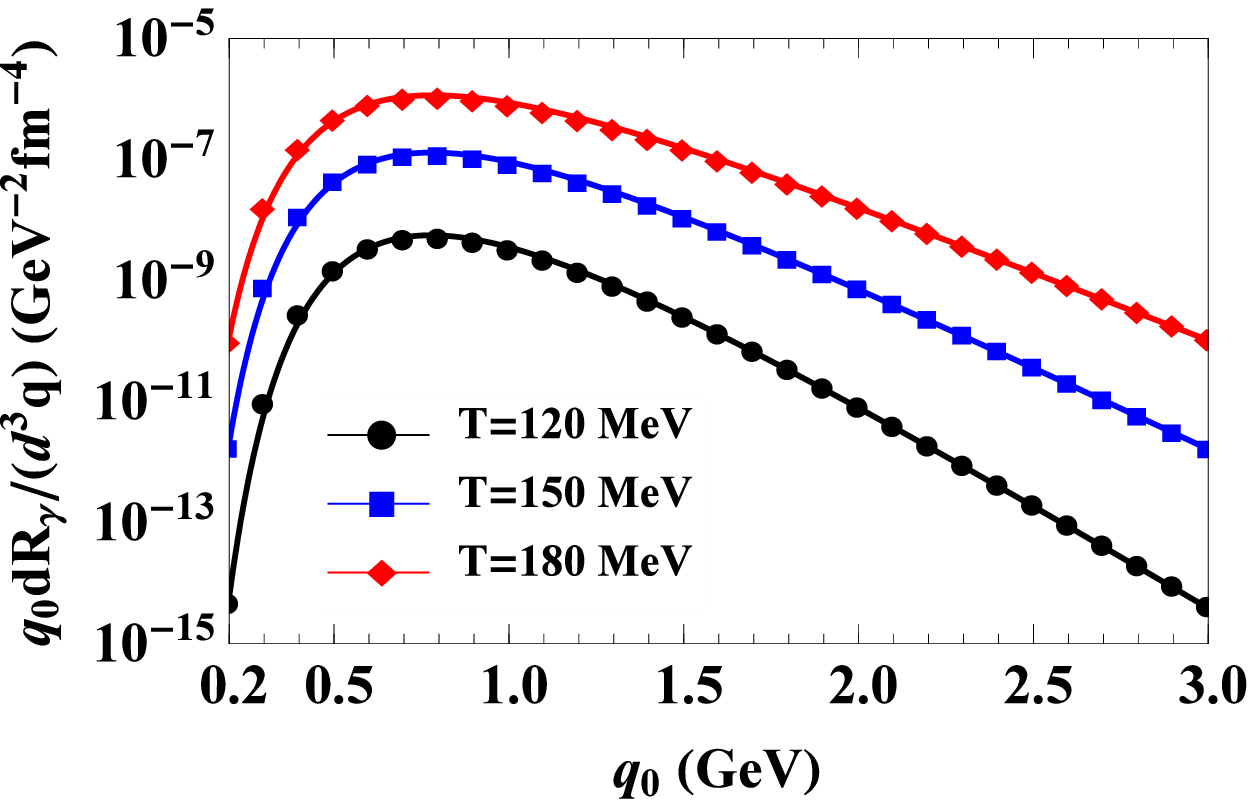}
\includegraphics[width=.435\textwidth]{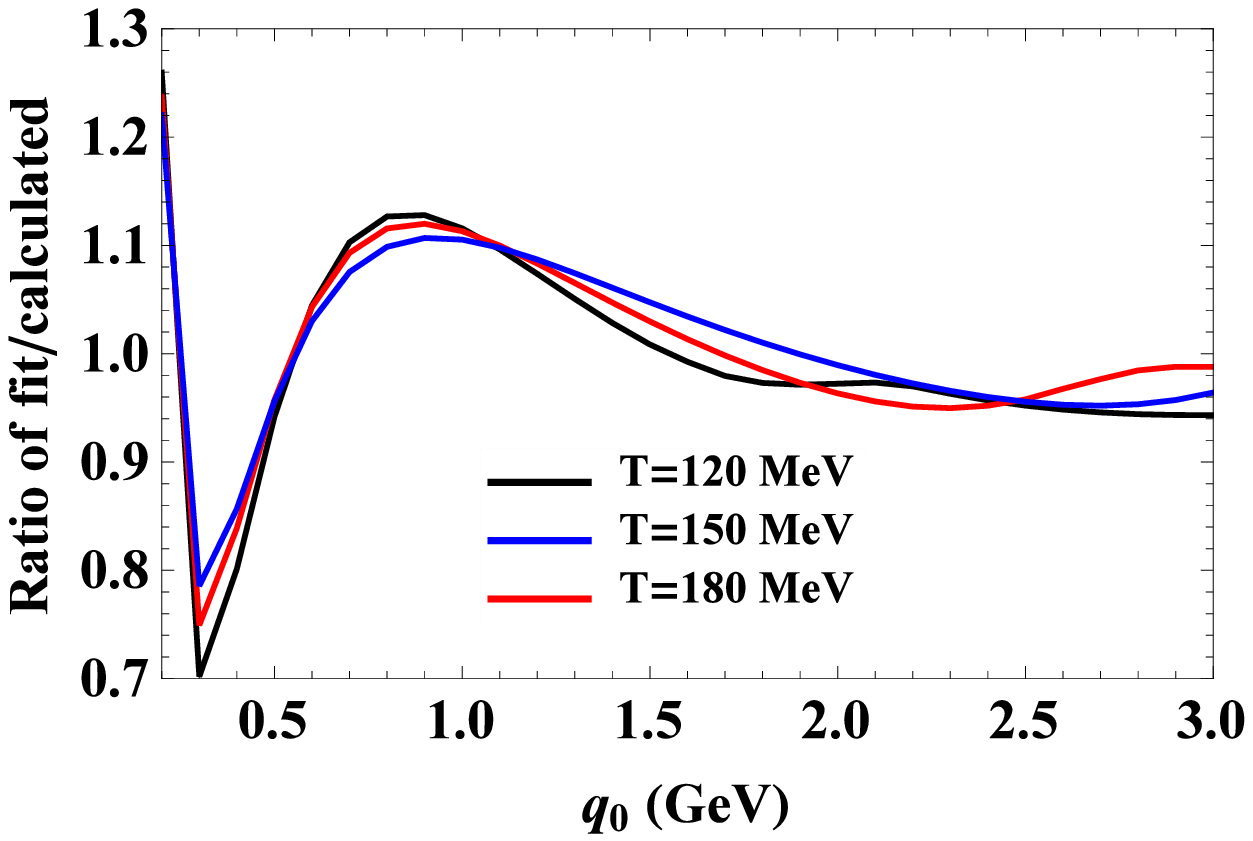}
\caption{Left panel: Calculated thermal photo-emission rates from the
$\rho \omega \to \gamma \pi$ process (symbols) compared to the parametrized
rates (lines). Right panel: Ratio of parametrized rates to
calculated rates.}
\end{figure}
\begin{eqnarray}
q_0 \frac{dR_{\rho \omega \to \gamma \pi}}{d^3q} & = & \exp
\big[ a_1q_0 + a_2 + \frac{a_3}{(q_0+0.2)} + \frac{a_4}{(q_0+0.2)^2} \big]
\, \left[\mathrm{fm}^{-4} \, \mathrm{GeV}^{-2} \right] \ , \\
 a_1(T) & = &  -29.6866 + 331.769 \, T - 1618.66 \,T^2 + 2918.53\, T^3 \ , \nonumber \\
 a_2(T) & = & -15.3332 + 90.2225 \,T - 300.185 \,T^2 + 428.386 \,T^3 \ ,  \\
 a_3(T) & = &  -7.35061 + 109.288 \,T - 630.396\, T^2 + 1227.69\, T^3 \ , \nonumber  \\
 a_4(T) & = & -10.6044 + 109.1 \,T - 500.718 \,T^2 + 872.951\, T^3 \ . \nonumber
 \end{eqnarray}

\twocolumngrid
\begin{flushleft}

\end{flushleft}
\end{document}